\documentclass[amsmath,amssymb,prb,superscriptaddress,twocolumn,showpacs,longbibliography]{revtex4-1}
\usepackage{graphicx}
\usepackage{dcolumn}
\usepackage[colorlinks=true,linkcolor=blue,citecolor=blue,urlcolor=blue]{hyperref}
\usepackage{stmaryrd}
\usepackage{esvect}
\usepackage{multirow}
\usepackage{upgreek}
\usepackage{soul}
\allowdisplaybreaks

\renewcommand{\vec}[1]{\boldsymbol{#1}}

\newcommand{\iu}{\mathrm{i}}
\newcommand{\Tr}{\mathrm{Tr}\,}

\newcommand{\MC}[1]{\mathcal{#1}}
\newcommand{\MR}[1]{\mathrm{#1}}
\newcommand{\VEC}[1]{\mathbf{#1}}

\usepackage{siunitx}
\DeclareSIUnit\mub{\mu_\text{B}}
\DeclareSIUnit\rydberg{\text{Ry}}
\usepackage{mathtools}

\DeclarePairedDelimiterX\braket[1]{\langle}{\rangle}{#1}

\usepackage{siunitx}
\DeclareSIUnit\mub{\mu_\text{B}}
\DeclareSIUnit\byte{\text{B}}
\DeclareSIUnit\rydberg{\text{Ry}}

\usepackage[usenames,dvipsnames]{xcolor}
\usepackage{tikz}
\usetikzlibrary{patterns}
\usetikzlibrary{decorations.markings}
\usetikzlibrary{arrows}
\usetikzlibrary{positioning}

\usepackage{changes}

\begin{document}

\tikzset{
    every node/.append style={font=\small},
    every edge/.append style={thick},
    arrow/.style={thick, shorten >=5pt,shorten <=5pt,->},
    Green_soc/.style={densely dashed, draw, postaction={decorate},
        decoration={markings,mark=at position .55 with {\arrow[draw]{>}}}},
    Green_full/.style={solid, double, draw, postaction={decorate},
        decoration={markings,mark=at position .55 with {\arrow[draw]{>}}}},
    Green/.style={solid, draw, postaction={decorate},
        decoration={markings,mark=at position .55 with {\arrow[draw]{>}}}},
    Green2/.style={solid, draw, postaction={decorate},
        decoration={markings,mark=at position .45 with {\arrow[draw]{>}}}},
    Green_soc2/.style={densely dashed, draw, postaction={decorate},
        decoration={markings,mark=at position .45 with {\arrow[draw]{>}}}},
    susc/.style={thick,fill=white,draw},
    dt/.style={thick,fill=lightgray,draw},
    soc/.style={thick,fill=black,draw},
    vertex/.style={thick,fill=black,draw}
}

\title{Prospecting chiral multi-site interactions in prototypical magnetic systems}

\author{Sascha Brinker}\email{s.brinker@fz-juelich.de}
\affiliation{Peter Gr\"{u}nberg Institut and Institute for Advanced Simulation, Forschungszentrum J\"{u}lich \& JARA, 52425 J\"{u}lich, Germany}
%Maybe next affiliation is not needed anymore.. but since I use some of the results from the thesis I will first leave it here
\affiliation{Department of Physics, RWTH Aachen University, 52056 Aachen, Germany} 
\author{Manuel dos Santos Dias}\email{m.dos.santos.dias@fz-juelich.de}
\affiliation{Peter Gr\"{u}nberg Institut and Institute for Advanced Simulation, Forschungszentrum J\"{u}lich \& JARA, 52425 J\"{u}lich, Germany}
\author{Samir Lounis}\email{s.lounis@fz-juelich.de}
\affiliation{Peter Gr\"{u}nberg Institut and Institute for Advanced Simulation, Forschungszentrum J\"{u}lich \& JARA, 52425 J\"{u}lich, Germany}

\date{\today}

\begin{abstract}
\noindent
Atomistic spin models have found enormous success in addressing the properties of magnetic materials, grounded on the identification of the relevant underlying magnetic interactions.
The huge development in the field of magnetic skyrmions and other noncollinear magnetic structures is largely due to our understanding of the chiral Dzyaloshinskii-Moriya interaction.
Recently, various works have proposed new types of chiral interactions, with seemingly different forms, but the big picture is still missing.
Here, we present a systematic construction of a generalized spin model containing isotropic and chiral multi-site interactions.
These are motivated by a microscopic model that incorporates local spin moments and the spin-orbit interaction, and their symmetry properties are established.
We show that the chiral interactions arise solely from the spin-orbit interaction and that the multi-site interactions do not have to follow Moriya’s rules, unlike the Dzyaloshinskii-Moriya and chiral biquadratic interactions.
The chiral multi-site interactions do not vanish due to inversion symmetry, and comply with a generalized Moriya rule: If all sites connected by the interaction lie in the same mirror plane, the chiral interaction vector must be perpendicular to this plane.
We then illustrate our theoretical considerations with density functional theory calculations for prototypical magnetic systems.
These are triangular trimers built out of Cr, Mn, Fe and Co adatoms on the Re(0001), Pt(111) and Au(111) surfaces, for which $C_\MR{3v}$ symmetry applies, and Cr and Fe square tetramers on Pt(001) with $C_\MR{4v}$ symmetry.
The multi-site interactions are substantial in magnitude and cannot be neglected when comparing the energy of different magnetic structures.
Finally, we discuss the recent literature in light of our findings, and clarify several unclear or confusing points.
\end{abstract}

\maketitle

\section{Introduction}
Atomistic spin models provide the foundation to understand the properties of magnetic materials: complex magnetic ground state structures, elementary excitations (spin waves), solitons whether topologically trivial or non-trivial (domain walls and magnetic skyrmions, respectively), thermal effects and real-time dynamics\cite{Eriksson2017}.
In comparison to the full quantum-mechanical description, this type of model aims at a coarse-grained, low-energy description of a given material, by assuming that magnetism is well-described by assigning rigid magnetic moments (spins) to specific spatial positions (sites), and specifying how these spins interact among each other (magnetic interactions).
Uncovering a new type of magnetic interaction often leads to novel magnetic states (such as quantum spin liquids\cite{Savary2017}), or even to new fields of research in magnetism (e.g.\ skyrmionics\cite{Fert2017}).
It is then essential to have a systematic catalogue of the possible magnetic interactions, preferably coupled with a theory that can make material-specific predictions and help guide or interpret experimental efforts.

Magnetic interactions can be broadly divided into isotropic and anisotropic interactions.
Isotropic interactions depend only on the relative angles between the spins, such as the original exchange interaction discovered by Heisenberg\cite{Heisenberg1928}, with detailed microscopic understanding provided for instance by Anderson's theory of superexchange\cite{Anderson1959}.
The anisotropic interactions depend on how the spins are aligned with real-space directions (for instance, the bonds between magnetic sites or certain crystal directions), and arise from relativistic effects, namely spin-orbit coupling (SOC).
Interactions which are symmetric under exchange of the spin components include the single-ion anisotropy and the two-site symmetric anisotropic exchange or compass anisotropy\cite{vanVleck1937,Moriya1953}, which together lead to the magnetocrystalline anisotropy.
An extreme case of symmetric anisotropic exchange is the Kitaev interaction\cite{Kitaev2006,Jackeli2009}, that can stabilized exotic quantum spin liquids\cite{Savary2017}.
A different kind of two-site anisotropic interaction, which is antisymmetric upon exchange of the spin components, is the Dzyaloshinskii-Moriya interaction (DMI)\cite{Dzyaloshinskii1958,Moriya1960}.
The DMI lifts the chiral degeneracy of the magnetic structure, as it favors one sense of rotation, leading to spiral magnetic ground states, domain walls and magnetic skyrmions of well-defined chirality or handedness.
All these interactions can also be classified by the number of spin components that they couple (2-spin interactions, sometimes more for the single-ion anisotropy) and by how many sites are coupled (1-site interaction for the single-ion anisotropy, the others being 2-site interactions).

There are two ways in which the previous set of interactions can be generalized, either by interactions that couple more spin components (without an external magnetic field: 4-spin, 6-spin, etc.) or more sites (3-site, 4-site, etc.), either isotropic or anisotropic.
The isotropic interactions include the biquadratic interaction (4-spin 2-site)\cite{Kittel1960,Harris1963,Huang1964}, the 4-spin 3-site interaction\cite{Uryu1965,Iwashita1974}, and the ring exchange (4-spin 4-site)\cite{Takahashi1977,MacDonald1988,MacDonald1990}, with a recent proposal for a 6-spin 3-site isotropic interaction\cite{Grytsiuk2020} which we will show can be related to the 6-spin 6-site interaction derived in Ref.~\onlinecite{MacDonald1990}.
These isotropic interactions can be systematically derived from a half-filled Hubbard model\cite{Takahashi1977,MacDonald1988,MacDonald1990,Hoffmann2020}.
They are rarely the dominant magnetic interactions, but can completely change the picture obtained from the previously-discussed 2-spin interactions alone \cite{Lounis2010,Szilva2013}.

The 2-spin interactions tend to favor relatively simple magnetic structures, such as ferromagnetic, antiferromagnetic, or spin spiral ground states --- these are called single-$\VEC{Q}$ states as their periodicity can be described by a single wavevector $\VEC{Q}$.
The isotropic 4-spin interactions can couple single-$\VEC{Q}$ states which are degenerate when considering only 2-spin interactions, and stabilize complex superpositions of such states.
The general mechanism and the possible connection to long-range interactions can be understood resorting to Kondo lattice models\cite{Batista2016,Ozawa2017,Hayami2017,Okumura2020}.
For example, the up-up-down-down state is a double-$\VEC{Q}$ state that was recently uncovered in magnetic monolayers with an hexagonal lattice\cite{Al-Zubi2011,Kroenlein2018,Romming2018}, with the 4-spin 3-site interaction playing the key role.
The nanoskyrmion lattice in an Fe monolayer on Ir(111)\cite{Heinze2011} can be seen as another type of double-$\VEC{Q}$ state stabilized by a combination of various interactions, including isotropic 4-spin ones.
The prediction of a triple-$\VEC{Q}$ state in this type of monolayers stabilized by the ring exchange\cite{Kurz2001} has also finally been experimentally realized\cite{Spethmann2020}.
Isotropic 4-spin interactions also explain why some bulk materials host short-period magnetic skyrmion lattices and other multiple-$\VEC{Q}$ states\cite{Tanigaki2015,Takagieaau2018,Fujishiro2019,Kurumaji2019,Khanh2020}.
Another class of materials where the isotropic 4-spin interactions play various roles are high-temperature superconductors, likely stabilizing the bicollinear antiferromagnetic ground state of FeTe\cite{Bao2009,Li2009} and modifying the spin wave spectrum of the parent compound La$_2$CuO$_4$\cite{Coldea2001}.
As a final example, solid $^3$He is perhaps the most famous system where multi-site isotropic interactions are essential to understand its magnetic properties\cite{Roger1983}, with up to 6-spin 6-site interactions quantitatively determined in an hexagonal monolayer\cite{Roger1998}.

In contrast to the large body of knowledge concerning isotropic multi-site interactions, not much attention has been paid to their anisotropic counterparts.
The anisotropic 2-spin interactions have been derived from an extended Hubbard model\cite{Yildirim1995}, but to our knowledge no attempt has been made to reach their 4-spin counterparts, so their possible forms remain unclear.
One can still proceed in various ways, which become very powerful if combined with a model-independent method of evaluating the energy of a magnetic structure, for instance resorting to density functional theory (DFT) calculations.
Firstly, the energy of different magnetic structures can be systematically mapped to a spin cluster expansion\cite{Drautz2004,Drautz2005,Antal2008,Singer2011,Szunyogh2011}, of which the four-state mapping method is a simplified form able to determine all types of 2-spin interactions (see for instance Ref.~\onlinecite{Xu2020} for an application to the Kitaev interaction).
In Ref.~\onlinecite{Brinker2019}, we used the spin-cluster expansion in combination with an intuitive microscopic model to catalogue all possible anisotropic 2-spin and 4-spin interactions in magnetic dimers on surfaces with strong SOC, uncovering the chiral biquadratic interaction (CBI) and also 3- and 4-site chiral interactions.
Phenomenological considerations can also be used to identify allowed forms for the interactions consistent with the symmetry of a target material. 
This led to the discovery of chiral 4-spin 3-site interactions in MnGe\cite{Grytsiuk2020} (for which a derivation based on multiple scattering theory was also provided), and motivated their existence in an Fe chain on Re(0001)\cite{Laszloffy2019}, while the magnetism of Ca$_3$Ru$_2$O$_7$ was rationalized by invoking higher-order Lifshitz invariants in connection to a Ginzburg-Landau theory\cite{Sokolov2019}.
Lastly, the energy can also be expanded in a Taylor series in small deviations from a reference magnetic structure.
This was developed into the very successful infinitesimal rotation method for 2-spin interactions\cite{Liechtenstein1987,Udvardi2003,Ebert2009}, with an extension to 4-spin interactions recently proposed\cite{Grytsiuk2020,Mankovsky2019}, and a growing body of work addressing its application for noncollinear magnetic structures\cite{Lounis2010,Szilva2013,Kvashnin2016,Cardias2020}.

In this work, we present a comprehensive study of isotropic and chiral multi-site interactions in prototypical magnetic systems.
First we specify the form of our spin model, containing besides 1-site and 2-site interactions (including the DMI and the CBI) also isotropic and chiral 3-site and 4-site interactions.
The employed forms of the interactions are fully justified by our microscopic model\cite{Brinker2019} (see Appendix B of that reference), and we supply simple heuristic arguments for their derivation.
We then discuss the symmetry properties of the chiral 3-site and 4-site interactions, showing in detail that they are not bound by Moriya's rules, in contrast to the DMI and the CBI.
After presenting our computational approach, we proceed to investigate these interactions in several magnetic systems.
We chose clusters that are common atomic motifs in many periodic magnetic materials, namely homoatomic trimers on fcc(111) and hcp(0001) surfaces ($C_\MR{3v}$ symmetry), and tetramers on the fcc(001) surface ($C_\MR{4v}$), which illustrate the magnitude and symmetry properties of the considered interactions.
Finally, we discuss several recent works in light of our findings, and present our conclusions.

\section{Spin model with isotropic and chiral multi-site interactions}
Consider a magnetic material with well-localized spin moments on atomic sites labelled $\{1,\ldots,N\}$.
The energy of a given spin configuration can be described by a function $\MC{E}(\VEC{S}_1,\ldots,\VEC{S}_N;\VEC{B})$ of the orientations of each spin moment, represented by classical unit vectors, $|\VEC{S}_i| = 1$, and of the external magnetic field $\VEC{B}$.
This energy function can contain several terms, describing different types of magnetic interactions, which we catalogue by how many spin components are involved ($p$-spin interactions), and by how many sites are involved ($q$-site interactions).
Time-reversal symmetry demands $\MC{E}(-\VEC{S}_1,\ldots,-\VEC{S}_N;-\VEC{B}) = \MC{E}(\VEC{S}_1,\ldots,\VEC{S}_N;\VEC{B})$.
This implies that the interactions which are independent of the external magnetic field must contain an even number of spin moments ($p \bmod 2 = 0$).
We now describe the interactions that we consider in this paper.
All the presented forms have already been derived from a microscopic model in our previous work, including the multi-site forms (see Appendix B of Ref.~\onlinecite{Brinker2019} for details).

\subsection{Spin model with 3- and 4-site interactions}
The magnetic interactions can be grouped in the following way, going from 1-site up to 4-site interactions:
\begin{equation}\label{eq:magen}
    \MC{E} = \sum_{i} \MC{E}_i + \frac{1}{2}\,{\sum_{i,j}}' \MC{E}_{ij} + \frac{1}{2}\,{\sum_{i,j,k}}' \MC{E}_{ijk}
    + \frac{1}{4}\,{\!\sum_{i,j,k,l}\!}' \MC{E}_{ijkl} \;.
\end{equation}
Here $i,j,k,l \in \{1,\ldots,N\}$ with $N$ the number of magnetic atoms.
The sums over sites are unrestricted except for the exclusion of repeated sites, $i \neq j \neq k \neq l$, which is indicated by the primes.
We now discuss the interactions that will be computed for the prototypical magnetic systems.

\subsection{1-site interactions}
The 1-site contribution to the magnetic energy is ($\alpha,\beta \in \{x,y,z\}$):
\begin{equation}\label{eq:1site}
    \MC{E}_1 = \VEC{B}\cdot\VEC{S}_1 + \sum_{\alpha,\beta} K_1^{\alpha\beta} S_1^\alpha S_1^\beta \;.
\end{equation}
We have a 1-spin interaction with the external magnetic field (in energy units), and the 2-spin interaction describes the lowest-order contribution to the on-site magnetic anisotropy energy.
As $S_1^\alpha S_1^\beta$ is a symmetric rank-2 tensor, only the symmetric part of $K_1^{\alpha\beta}$ contributes to the sum, so it can be written as
\begin{equation}\label{eq:kmatrix}
    \renewcommand{\arraystretch}{1.2}
    K_1 = \begin{pmatrix} K_1^{xx} & K_1^{xy} & K_1^{xz} \\ K_1^{xy} & K_1^{yy} & K_1^{yz} \\  K_1^{xz} & K_1^{yz} & K_1^{zz} \end{pmatrix} \;.
\end{equation}
Its eigenvectors specify the easy, intermediate and hard local anisotropy axes, in increasing order of the corresponding energy eigenvalues.
The non-vanishing elements of this matrix are determined by the local symmetry of the environment of the magnetic atom.
Furthermore, the condition $|\VEC{S}_1| = 1$ removes one parameter, so five independent parameters remain.
The choice of free parameters used in this work is explained in Appendix~\ref{app:wizardry}.

\subsection{2-site interactions}
For the 2-site interactions we consider all possible 2-spin interactions plus two kinds of 4-spin interactions:
\begin{align}\label{eq:2site}
    \MC{E}_{12} &= J_{12}\,\VEC{S}_1\cdot\VEC{S}_2 + \VEC{D}_{12}\cdot\left(\VEC{S}_1 \times \VEC{S}_2\right) + \sum_{\alpha,\beta} \Delta J_{12}^{\alpha\beta} S_1^\alpha S_2^\beta  \nonumber\\
    &+ B_{12} \left(\VEC{S}_1\cdot\VEC{S}_2\right)^2
    + \VEC{C}_{12} \cdot \left(\VEC{S}_1\times\VEC{S}_2\right) \left(\VEC{S}_1\cdot\VEC{S}_2\right) \;.
\end{align}
Here $J_{12}$ is the conventional isotropic Heisenberg exchange interaction\cite{Heisenberg1928}, $\VEC{D}_{12}$ is the vector defining the chiral Dzyaloshinskii-Moriya interaction\cite{Dzyaloshinskii1958,Moriya1960} which can only be present in systems without inversion symmetry, and $\Delta J_{12}^{\alpha\beta} = \Delta J_{12}^{\beta\alpha}$ defines the symmetric exchange anisotropy\cite{Hermenau2019}.
Like the on-site magnetic anisotropy, the eigenvectors of the $\Delta J_{12}$ matrix can be used to specify the easy, intermediate and hard axes of a given pair of magnetic atoms (see Appendix~\ref{app:wizardry}).
For very anisotropic systems this leads to the Kitaev bond-dependent symmetric exchange anisotropy\cite{Jackeli2009,Xu2020}.
The 2-site interactions are augmented with the isotropic biquadratic interaction $B_{12} \left(\VEC{S}_1\cdot\VEC{S}_2\right)^2$ and with the recently-uncovered chiral biquadratic interaction $\VEC{C}_{12} \cdot \left(\VEC{S}_1\times\VEC{S}_2\right) \left(\VEC{S}_1\cdot\VEC{S}_2\right)$, which follows the same Moriya rules as the 2-spin DMI.
These were found to be the dominant 4-spin contributions to the 2-site interactions in our previous study, Ref.~\onlinecite{Brinker2019}.
If the relativistic spin-orbit interaction can be neglected then only the isotropic Heisenberg and biquadratic interactions remain.
The interaction coefficients have the general symmetries $J_{12} = J_{21}$, $\VEC{D}_{12} = -\VEC{D}_{21}$, $\Delta J_{12}^{\alpha\beta} = \Delta J_{21}^{\alpha\beta}$, $B_{12} = B_{21}$ and $\VEC{C}_{12} = -\VEC{C}_{21}$, which justify the prefactor of $1/2$ assigned to the 2-site interactions in Eq.~\eqref{eq:magen}.

\subsection{Heuristic arguments for the form of the chiral multi-site interactions}
Instead of repeating the derivations already presented in Ref.~\onlinecite{Brinker2019}, here we present heuristic arguments for the construction of chiral multi-site interactions starting from known isotropic multi-site interactions.
The microscopic model discussed in that work presents a systematic expansion of the grand potential in terms of the spin-dependent parts of the electronic hamiltonian.
These are assumed to be a local exchange coupling between the electrons and the local moments $\propto \vec{\upsigma} \cdot \VEC{S}_i$, and the local SOC $\propto \vec{\upsigma} \cdot \VEC{L}$ (its spatial location can be left unspecified), with $\vec{\upsigma}$ the vector of Pauli matrices describing the electron spin, $\VEC{S}_i$ a unit vector describing the orientation of the spin moment at site $i$, and $\VEC{L}$ the local orbital angular momentum operator.
Isotropic interactions are represented by closed loops that contain an even number of magnetic sites and are connected by spin-independent Green functions, while chiral interactions additionally contain one spin-orbit site.
The specifics of the diagrammatic expansion of the grand potential are not needed if one is only interested in the form of the interactions, and it is based this point of view that we now present our heuristic arguments.
What is essential is in which order the magnetic and SOC sites appear, and the algebra of Pauli matrices.
The final goal is to identify the elementary forms of the interactions which are needed to construct the atomistic spin model according to Eq.~\eqref{eq:magen}.

    \begin{table}[!t]
        \centering
        \begin{ruledtabular}
            \begin{tabular}{lccc}
            $p$-spin $q$-site & isotropic & & chiral \\\hline
            2-spin 2-site &
                    \begin{tikzpicture}[baseline={([yshift=-.5ex]current bounding box.center)}]
                      \begin{scope}[scale=2]
                          % coordinates
                          \coordinate (lb) at (0,0);
                          \coordinate (mt) at (0.25,-0.25);
                          \coordinate (rb) at (0.5,0);
                          \coordinate (lt) at (0,0.5);
                          \coordinate (mt) at (0.25,0.25);
                          \coordinate (rt) at (0.5,0.5);
                          % edges
                          \draw (lb) edge[Green,out=-45,in=-135] node[below,yshift=-.2cm] {} (rb);
                          \draw (rb) edge[Green,out=135,in=45] node[above,yshift=.2cm] {} (lb);
                          % arrows
                          \draw[susc] (lb) circle (1pt) node[left] {$1 \,$};
                          \draw[susc] (rb) circle (1pt) node[right] {$\, 2 $};
                      \end{scope}
                    \end{tikzpicture} 
                    & $\longrightarrow$ &
                    \begin{tikzpicture}[baseline={([yshift=-.5ex]current bounding box.center)}]
                      \begin{scope}[scale=2]
                          % coordinates
                          \coordinate (lb) at (0,0);
                          \coordinate (rb) at (0.5,0);
                          % edges
                          \draw (lb) edge[Green_soc,out=-45,in=-135] node[below,yshift=-.2cm] {} (rb);
                          \draw (rb) edge[Green,out=135,in=45] node[above,yshift=.2cm] {} (lb);
                          % arrows
                          \draw[susc] (lb) circle (1pt) node[left] {$1 \,$};
                          \draw[susc] (rb) circle (1pt) node[right] {$\, 2 $};
                      \end{scope}
                    \end{tikzpicture}\\
            4-spin 2-site & 
                    \begin{tikzpicture}[baseline={([yshift=-.5ex]current bounding box.center)}]
                      \begin{scope}[scale=2]
                          % coordinates
                          \coordinate (lb) at (0,0);
                          \coordinate (rb) at (0.5,0);
                          % edges
                          \draw (lb) edge[Green,out=-25,in=-155] node[below,yshift=-.2cm] {} (rb);
                          \draw (rb) edge[Green,out=155,in=25] node[above,yshift=.2cm] {} (lb);
                          \draw (lb) edge[Green,out=-80,in=-100] node[below,yshift=-.2cm] {} (rb);
                          \draw (rb) edge[Green,out=100,in=80] node[above,yshift=.2cm] {} (lb);
                          % arrows
                          \draw[susc] (lb) circle (1pt) node[left] {$1 \,$};
                          \draw[susc] (rb) circle (1pt) node[right] {$\, 2$};
                      \end{scope}
                    \end{tikzpicture}
                    & $\longrightarrow$ & 
                    \begin{tikzpicture}[baseline={([yshift=-.5ex]current bounding box.center)}]
                      \begin{scope}[scale=2]
                          % coordinates
                          \coordinate (lb) at (0,0);
                          \coordinate (rb) at (0.5,0);
                          % edges
                          \draw (lb) edge[Green,out=-25,in=-155] node[below,yshift=-.2cm] {} (rb);
                          \draw (rb) edge[Green,out=155,in=25] node[above,yshift=.2cm] {} (lb);
                          \draw (lb) edge[Green_soc,out=-80,in=-100] node[below,yshift=-.2cm] {} (rb);
                          \draw (rb) edge[Green,out=100,in=80] node[above,yshift=.2cm] {} (lb);
                          % arrows
                          \draw[susc] (lb) circle (1pt) node[left] {$1 \,$};
                          \draw[susc] (rb) circle (1pt) node[right] {$\, 2$};
                      \end{scope}
                    \end{tikzpicture}
                    \\
            4-spin 3-site & 
                    \begin{tikzpicture}[baseline={([yshift=-.5ex]current bounding box.center)}]
                      \begin{scope}[scale=2]
                          % coordinates
                          % coordinates
                          \coordinate (mb) at (0,0);
                          \coordinate (lb) at (-.5,0);
                          \coordinate (rb) at (.5,0);
                          % edges
                          \draw (lb) edge[Green,out=-45,in=-135] node[below,yshift=-.2cm] {} (mb);
                          \draw (mb) edge[Green,out=135,in=45] node[above,yshift=.2cm] {} (lb);
                          \draw (mb) edge[Green,out=-45,in=-135] node[below,yshift=-.2cm] {} (rb);
                          \draw (rb) edge[Green,out=135,in=45] node[above,yshift=.2cm] {} (mb);
                          % arrows
                          \draw[susc] (mb) circle (1pt) node[above,yshift=.2cm] {$2 $};
                          \draw[susc] (lb) circle (1pt) node[left] {$1 \, $};
                          \draw[susc] (rb) circle (1pt) node[right] {$\, 3 $};
                      \end{scope}
                    \end{tikzpicture}
                    & $\longrightarrow$ & 
                    \begin{tikzpicture}[baseline={([yshift=-.5ex]current bounding box.center)}]
                      \begin{scope}[scale=2]
                          % coordinates
                          % coordinates
                          \coordinate (mb) at (0,0);
                          \coordinate (lb) at (-.5,0);
                          \coordinate (rb) at (.5,0);
                          % edges
                          \draw (lb) edge[Green_soc,out=-45,in=-135] node[below,yshift=-.2cm] {} (mb);
                          \draw (mb) edge[Green,out=135,in=45] node[above,yshift=.2cm] {} (lb);
                          \draw (mb) edge[Green,out=-45,in=-135] node[below,yshift=-.2cm] {} (rb);
                          \draw (rb) edge[Green,out=135,in=45] node[above,yshift=.2cm] {} (mb);
                          % arrows
                          \draw[susc] (mb) circle (1pt) node[above,yshift=.2cm] {$2 $};
                          \draw[susc] (lb) circle (1pt) node[left] {$1 \, $};
                          \draw[susc] (rb) circle (1pt) node[right] {$\, 3 $};
                      \end{scope}
                    \end{tikzpicture}
                    \\
            4-spin 4-site & 
                    \begin{tikzpicture}[baseline={([yshift=-.5ex]current bounding box.center)}]
                      \begin{scope}[scale=2]
                          % coordinates
                          \coordinate (lb) at (0,0);
                          \coordinate (rb) at (0.5,0);
                          \coordinate (rt) at (0.5,0.5);
                          \coordinate (lt) at (0.0,0.5);
                          % edges
                          \draw (lb) edge[Green,out=-45,in=-135] node[left,xshift=-.2cm] {} (rb);
                          \draw (rb) edge[Green,out=45,in=-45] node[right,xshift=.2cm] {} (rt);
                          \draw (rt) edge[Green,out=135,in=45] node[right,yshift=.2cm] {} (lt);
                          \draw (lt) edge[Green,out=-135,in=135] node[left,yshift=.2cm] {} (lb);
                          % arrows
                          \draw[susc] (lb) circle (1pt) node[left,yshift=-.3cm] {$1$};
                          \draw[susc] (rb) circle (1pt) node[right,yshift=-.3cm] {$2$};
                          \draw[susc] (rt) circle (1pt) node[right,yshift=.3cm] {$3$};
                          \draw[susc] (lt) circle (1pt) node[left,yshift=.3cm] {$4$};
                      \end{scope}
                    \end{tikzpicture}
                    & $\longrightarrow$ &
                    \begin{tikzpicture}[baseline={([yshift=-.5ex]current bounding box.center)}]
                      \begin{scope}[scale=2]
                          % coordinates
                          \coordinate (lb) at (0,0);
                          \coordinate (rb) at (0.5,0);
                          \coordinate (rt) at (0.5,0.5);
                          \coordinate (lt) at (0.0,0.5);
                          % edges
                          \draw (lb) edge[Green_soc,out=-45,in=-135] node[left,xshift=-.2cm] {} (rb);
                          \draw (rb) edge[Green,out=45,in=-45] node[right,xshift=.2cm] {} (rt);
                          \draw (rt) edge[Green,out=135,in=45] node[right,yshift=.2cm] {} (lt);
                          \draw (lt) edge[Green,out=-135,in=135] node[left,yshift=.2cm] {} (lb);
                          % arrows
                          \draw[susc] (lb) circle (1pt) node[left,yshift=-.3cm] {$1$};
                          \draw[susc] (rb) circle (1pt) node[right,yshift=-.3cm] {$2$};
                          \draw[susc] (rt) circle (1pt) node[right,yshift=.3cm] {$3$};
                          \draw[susc] (lt) circle (1pt) node[left,yshift=.3cm] {$4$};
                      \end{scope}
                    \end{tikzpicture}
                    \\
            \end{tabular}
        \end{ruledtabular}
    	\caption{
    	Diagrams for the heuristic derivation of the form of the magnetic interactions.
    	The numbered sites represent local exchange couplings of the type $\vec{\upsigma} \cdot \VEC{S}_i$, and the interaction is given by a closed loop connecting the sites with an even number of lines.
    	The form of the isotropic interaction can then be obtained by a spin trace over the ordered product of sites, according to the order in which the loop is traveled.
    	The form of the chiral interactions (first-order in SOC) can be obtained from the diagrams for the isotropic interactions by inserting $\vec{\upsigma} \cdot \VEC{L}$ in-between a pair of sites (dashed line).
    	As a set of sites can be connected in various ways, this will lead to various forms for the interactions, which can be related to each other via symmetry operations, if applicable.}
    	\label{tab:Diagrams}
    \end{table}

As shown in Table~\ref{tab:Diagrams}, the isotropic 2-site interactions arise from the loops connecting 1 and 2 once (2-spin, bilinear) and twice (4-spin, biquadratic).
The form of these interactions can be obtained by tracing a product of local exchange couplings, according to the order that the sites are travelled in the loop.
For the isotropic bilinear interaction one then writes %(the prefactor is only for convenience)
\begin{align}
    &\frac{1}{2}\,\Tr (\vec{\upsigma} \cdot \VEC{S}_1) (\vec{\upsigma} \cdot \VEC{S}_2) 
    = \VEC{S}_1\cdot \VEC{S}_2 \;,
\end{align}
and for the biquadratic one
\begin{align}
    \frac{1}{2}\,\Tr (\vec{\upsigma} \cdot \VEC{S}_1) (\vec{\upsigma} \cdot \VEC{S}_2) (\vec{\upsigma} \cdot \VEC{S}_1) (\vec{\upsigma} \cdot \VEC{S}_2)
    = 2 \left(\VEC{S}_1\cdot \VEC{S}_2\right)^2 - 1 \;.
\end{align}
The second term is a constant and can be disregarded.
For these interactions it clearly does not matter if the loop starts from site $1$ or from site $2$.

The corresponding chiral 2-site interactions can be generated by inserting spin-orbit coupling in-between two magnetic sites.
For the bilinear interaction, inserting SOC between 1 and 2 gives (the connection containing SOC is indicated by the dashed line)
\begin{align}
    \frac{1}{2\iu}\,\Tr (\vec{\upsigma} \cdot \VEC{S}_1) (\vec{\upsigma} \cdot \VEC{L}) (\vec{\upsigma} \cdot \VEC{S}_2)
    = \VEC{L}\cdot\left(\VEC{S}_2 \times \VEC{S}_1\right) \;.
\end{align}
This generates the DMI.
Similarly, starting from the biquadratic interaction and inserting SOC between $1$ and $2$ produces the chiral biquadratic coupling $\VEC{S}_2 \times \VEC{S}_1 \left(\VEC{S}_1 \cdot \VEC{S}_2\right)$, with similar properties to the DMI.
The chiral interaction vectors are then governed by the SOC, and it can also be shown that they comply with the symmetry operations relating $1$ and $2$ (see Ref.~\onlinecite{Brinker2019}).
In the following we will present the simplified forms of the isotropic and chiral 3-site and 4-site interactions, along with the heuristic arguments that justify them.

\subsection{3-site interactions}
As with the 4-spin 2-site interactions, here we will also restrict our attention to isotropic and chiral interactions.
Due to time-reversal symmetry, a 3-site interaction must contain an even number of spin moments, 4-spin being the minimum, which implies that at least one of them appears repeatedly.
We adopt the convention of Laszloffy \textit{et al.}~\cite{Laszloffy2019} that the second site is the one that will appear repeated, see Table~\ref{tab:Diagrams}.
The isotropic 3-site interaction is thus expected to have the form
\begin{align}
    &\frac{1}{2}\,\Tr (\vec{\upsigma} \cdot \VEC{S}_1) (\vec{\upsigma} \cdot \VEC{S}_2) (\vec{\upsigma} \cdot \VEC{S}_3) (\vec{\upsigma} \cdot \VEC{S}_2) \nonumber\\
    &= 2\left(\VEC{S}_1\cdot \VEC{S}_2\right) \left(\VEC{S}_2\cdot \VEC{S}_3\right) - \VEC{S}_1\cdot\VEC{S}_3 \;.
\end{align}
The second term is of the form of the isotropic 2-site interaction and so it can be dropped, leaving the first term as our prototype for isotropic 4-spin 3-site interactions.
The basic symmetry is that $(1,2,3)$ and $(3,2,1)$ produce the same interaction.

If we insert SOC between 1 and 2, we obtain
\begin{align}
    &\frac{1}{2\iu}\,\Tr (\vec{\upsigma} \cdot \VEC{S}_1) (\vec{\upsigma} \cdot \VEC{L}) (\vec{\upsigma} \cdot \VEC{S}_2) (\vec{\upsigma} \cdot \VEC{S}_3) (\vec{\upsigma} \cdot \VEC{S}_2) \nonumber\\
    &= 2\,\VEC{L}\cdot\left(\VEC{S}_2 \times \VEC{S}_1\right) \left(\VEC{S}_2\cdot \VEC{S}_3\right)
    + \VEC{L}\cdot\left(\VEC{S}_1\times\VEC{S}_3\right) \;. \label{eq:chiral_3-site_diagram}
\end{align}
The second term is of the form of the DMI and so it can be dropped, leaving the first term as our prototype for chiral 4-spin 3-site interactions.
There is no relation between $(1,2,3)$ and $(3,2,1)$, due to the cross product in the first term.
This can be understood from our heuristic argument: SOC can influence different pairs of sites, which can naturally lead to different chiral interaction vectors, depending on the symmetry of the system.
If one had instead inserted SOC between 1 and 3 one would find a similar form, but with the dot product now between 1 and 2 and the cross product between 3 and 2.
Inserting SOC in other places does not bring other forms for the spin coupling, up to minus signs.

We then write the 3-site interactions as
\begin{align}\label{eq:3site}
    \MC{E}_{123} &= \, B_{123} \left(\VEC{S}_1\cdot\VEC{S}_2\right) \left(\VEC{S}_2\cdot\VEC{S}_3\right) \nonumber\\
    &+ \VEC{C}_{123} \cdot \left(\VEC{S}_1\times \VEC{S}_2\right) \left(\VEC{S}_2\cdot\VEC{S}_3\right) \;.
\end{align}
Note the convention that the second site is the one that appears repeated.
The other possible combination of dot and cross products is covered by $\MC{E}_{321}$, which is included in Eq.~\eqref{eq:magen}.
The isotropic interaction has the general symmetry $B_{123} = B_{321}$, which justifies the prefactor of $1/2$ in Eq.~\eqref{eq:magen}.
An alternative way of expressing the chiral interactions is using the symmetric and antisymmmetric combinations:
\begin{equation}
    \VEC{C}_{123}^\pm \cdot \big(\VEC{S}_1\times \VEC{S}_2 \left(\VEC{S}_2\cdot\VEC{S}_3\right)
    \pm {\VEC{S}_3\times \VEC{S}_2 \left(\VEC{S}_2\cdot\VEC{S}_1\right)}\big) \;,
\end{equation}
and the corresponding chiral interaction vectors have the general symmetry $\VEC{C}_{123}^\pm = \pm\VEC{C}_{321}^\pm$.
The minus combination can be expressed using the scalar spin chirality (see Appendix~\ref{app:sci}), providing a link to the spin-chiral interactions introduced in Ref.~\onlinecite{Grytsiuk2020}:
\begin{align}
    &\VEC{C}_{123}^- \cdot \left(\VEC{S}_1\times \VEC{S}_2 \left(\VEC{S}_2\cdot\VEC{S}_3\right)
    - \VEC{S}_3\times \VEC{S}_2 \left(\VEC{S}_2\cdot\VEC{S}_1\right)\right) \\
    &= \left(\VEC{C}_{123}^- \cdot \VEC{S}_2\right) \VEC{S}_1\cdot\left(\VEC{S}_2\times\VEC{S}_3\right) + \VEC{C}_{123}^-\cdot\left(\VEC{S}_1\times\VEC{S}_3\right) \;.
\end{align}

\subsection{4-site interactions}
Following our heuristic argument, the form of the isotropic 4-site interactions can obtained from a 4-site loop as shown in Table~\ref{tab:Diagrams}, which translates to
\begin{align}
    &\frac{1}{2}\,\Tr (\vec{\upsigma} \cdot \VEC{S}_1) (\vec{\upsigma} \cdot \VEC{S}_2) (\vec{\upsigma} \cdot \VEC{S}_3) (\vec{\upsigma} \cdot \VEC{S}_4) \nonumber\\
    &= \left(\VEC{S}_1\cdot \VEC{S}_2\right) \left(\VEC{S}_3\cdot \VEC{S}_4\right)
    - \left(\VEC{S}_1\cdot \VEC{S}_3\right) \left(\VEC{S}_2\cdot \VEC{S}_4\right) \nonumber\\
    &+ \left(\VEC{S}_1\cdot \VEC{S}_4\right) \left(\VEC{S}_2\cdot \VEC{S}_3\right)\;.
\end{align}
This is the known form of the ring exchange, including the minus sign.
We take just the first term as our prototype for isotropic 4-spin 4-site interactions, as the site summations in Eq.~\eqref{eq:magen} will reproduce the remaining possibilities for combining pairs of sites with dot products, and the interaction coefficients will cover the symmetry (see e.g.\ Ref.~\onlinecite{Hoffmann2020}).

The form of the chiral interaction can be obtained as before, by inserting SOC between 1 and 2:
\begin{align}
    &\frac{1}{2\iu}\,\Tr (\vec{\upsigma} \cdot \VEC{S}_1) (\vec{\upsigma} \cdot \VEC{L}) (\vec{\upsigma} \cdot \VEC{S}_2) (\vec{\upsigma} \cdot \VEC{S}_3) (\vec{\upsigma} \cdot \VEC{S}_4) \nonumber\\
    &= \VEC{L} \cdot \left(\VEC{S}_2 \times \VEC{S}_1\right) \left(\VEC{S}_3 \cdot \VEC{S}_4\right)
    + \VEC{L} \cdot \left(\VEC{S}_3 \times \VEC{S}_4\right) \left(\VEC{S}_1\cdot \VEC{S}_2\right) \nonumber\\
    &+ \VEC{L} \cdot \left(\VEC{S}_1 \times \VEC{S}_3\right) \left(\VEC{S}_2 \cdot \VEC{S}_4\right)
    + \VEC{L} \cdot \left(\VEC{S}_4 \times \VEC{S}_2\right) \left(\VEC{S}_1 \cdot \VEC{S}_3\right) \nonumber\\
    &+ \VEC{L} \cdot \left(\VEC{S}_2 \times \VEC{S}_3\right) \left(\VEC{S}_1 \cdot \VEC{S}_4\right)
    + \VEC{L} \cdot \left(\VEC{S}_4 \times \VEC{S}_1\right) \left(\VEC{S}_2 \cdot \VEC{S}_3\right) \;. \label{eq:microscopic_model_chiral_4-site}
\end{align}
This can be obtained from the form of the ring exchange by replacing a dot product by a cross product in every term, and leaving the other dot product, with an additional minus sign if the second dot product is replaced.
The perhaps unexpected complexity of this interaction can be understood from the corresponding diagram.
In contrast to the chiral 3-site interaction, where SOC only affects one of two bubbles in the diagram, here SOC affects the entire loop, even if it is inserted between a specific pair of sites, and thus generates all possible kinds of pairwise chiral couplings between the four spins.
As for the chiral 3-site interaction, inserting SOC between a different pair of sites leads to a different form with an independent chiral interaction vector.
Once again, by exploiting the summation over sites in Eq.~\eqref{eq:magen} it is sufficient to take the first term as our prototype for the chiral 4-spin 4-site interaction.

We thus express the contribution to the magnetic energy from the isotropic and chiral 4-site interactions as
\begin{align}\label{eq:4site}
    \MC{E}_{1234} &= B_{1234}\,(\VEC{S}_1\cdot\VEC{S}_2) (\VEC{S}_3\cdot\VEC{S}_4) \nonumber\\
    &+ \VEC{C}_{1234} \cdot \left(\VEC{S}_1\times\VEC{S}_2\right) \left( \VEC{S}_3\cdot \VEC{S}_4\right) \;.
\end{align}
Note the convention in these interactions that the sites are paired as (1,2) and (3,4), and that the cross product applies to the first pair.
If we repeat one site and write (1,2,2,3) this form reduces to the 3-site one.
We also have the general symmetries for the interaction coefficients $B_{1234} = B_{2134} = B_{1243} = B_{2143}$ and $\VEC{C}_{1234} = \VEC{C}_{1243} = -\VEC{C}_{2134} = -\VEC{C}_{2143}$, which justify the prefactor of $1/4$ assigned to the 4-site interactions in Eq.~\eqref{eq:magen}.

\section{Symmetries of multi-site interactions}
Following Neumann's principle, the magnetic energy function must respect the point group symmetry.
If $\MC{G}$ is a symmetry operation of the point group, this means $\MC{E}(\MC{G}\VEC{S}_1,\ldots,\MC{G}\VEC{S}_N;\MC{G}\VEC{B}) = \MC{E}(\VEC{S}_1,\ldots,\VEC{S}_N;\VEC{B})$.
In the absence of an external magnetic field, all interactions must then be invariant under all symmetry operations of the \emph{crystallographic} point group.
The action of the symmetry operation can be separated as $\MC{G} = \MC{P} \MC{O}$, where $\MC{P}$ maps the atomic sites to each other, and $\MC{O}$ transforms the orientations of the magnetic moments (rotation $\MC{R}$, mirroring $\MC{M}$, inversion $\MC{I}$).
The matrices $\MC{O}$ are orthogonal, $\MC{O}^{-1} = \MC{O}^\MR{T}$, and in particular for the mirror symmetries $\MC{M} = \MC{M}^\MR{T}$.
For a magnetic interaction connecting $q$-sites $(i_1,\ldots,i_q)$, a symmetry operation that maps these sites into themselves, $\MC{P}(i_1,\ldots,i_q) = (i_1,\ldots,i_q)$, can place constraints on the interaction coefficients (e.g.\ Moriya's rules).
If this is not the case, i.e.\ $\MC{P}(i_1,\ldots,i_q) \neq (i_1,\ldots,i_q)$, then we only find relations between $q$-site interactions connecting different sets of sites.

The basic building blocks of the interactions that we discuss in this work are either dot products $\VEC{S}_i\cdot\VEC{S}_j$ or cross products  $\VEC{S}_i\times\VEC{S}_j$ of the spin orientations, which are combined in various ways for the different types of interactions.
A symmetry operation $\MC{G}$ acts on these building blocks as follows.
The relation between atomic sites implied by the symmetry operation is expressed by the replacement $\MC{P}(i,j) = (k,l)$.
The dot product transforms as $(\MC{G}\VEC{S}_i)\cdot(\MC{G}\VEC{S}_j) = (\MC{O}\VEC{S}_k)\cdot(\MC{O}\VEC{S}_l) = \VEC{S}_k\cdot\VEC{S}_l$, and the last equality follows from the fact that the spatial transformations $\MC{O}$ leave the angle between vectors unchanged. 
If the interaction consists solely of dot products of spin orientations then the symmetry operations establish relations between the interaction coefficients, e.g.\ $J_{ij}\,(\MC{G}\VEC{S}_i)\cdot(\MC{G}\VEC{S}_j) = J_{ij}\,\VEC{S}_k\cdot\VEC{S}_l$, which implies $J_{ij} = J_{kl}$.
For the cross product there is a subtlety: $(\MC{G}\VEC{S}_i)\times(\MC{G}\VEC{S}_j) = (\MC{O}\VEC{S}_k)\times(\MC{O}\VEC{S}_l) = \left(\det \MC{O}\right) \MC{O}(\VEC{S}_k\times\VEC{S}_l)$.
As the cross product is an axial vector, it transforms for proper rotations with $\det \MC{O} = +1$ and for improper rotations (inversion, mirroring) with $\det \MC{O} = -1$.
For the chiral interactions, the cross product is combined with the chiral interaction vector, for instance $\VEC{D}_{ij}$ for the DMI, which then transfers the result of the spatial symmetry from the cross product to this vector, $\VEC{D}_{ij}\cdot\big((\MC{G}\VEC{S}_i)\times(\MC{G}\VEC{S}_j)\big) = \left(\det \MC{O}\right) (\MC{O}^{-1}\VEC{D}_{ij})\cdot(\VEC{S}_k\times\VEC{S}_l)$.
This implies the relation $\VEC{D}_{kl} = \left(\det \MC{O}\right) (\MC{O}^{-1}\VEC{D}_{ij})$.
If $(k,l) = (i,j)$ up to reordering we find constraints on the allowed components of the axial vector, which leads to Moriya's rules\cite{Moriya1960}.
For instance, $\{\VEC{S}_i,\VEC{S}_j\} \rightarrow \{\VEC{S}_j,\VEC{S}_i\}$ if an inversion center is present between $i$ and $j$, which leads to $\VEC{D}_{ij} = -\VEC{D}_{ij}$ and confirms the vanishing of the DMI (and CBI) in this case.
If $(k,l) \neq (i,j)$, we find relations between chiral vectors connecting different sites, and this shows that the vectors must be related by a simple change in orientation, as the spatial symmetry $\MC{O}$ leaves the length of vectors invariant.
The properties of the different multi-site interactions then follow from combining these principles with each type of interaction and the symmetry of the considered system.

For the benefit of the reader, we briefly recall the Moriya rules\cite{Moriya1960} for the DMI vector acting on the bond between sites $i$ and $j$:
\begin{enumerate}
    \item If there is an inversion center in the middle of the bond, $\VEC{D}_{ij} = 0$.
    \item If the bond is bisected by a mirror plane, $\VEC{D}_{ij}$ must lie in this plane.
    \item If the bond is contained in a mirror plane, $\VEC{D}_{ij}$ must be perpendicular to this plane.
    \item If a twofold rotation axis passes through the middle of the bond, $\VEC{D}_{ij}$ must be perpendicular to this axis.
    \item If the bond lies on an $n$-fold rotation axis,  $\VEC{D}_{ij}$ must be along this axis.
\end{enumerate}
They follow from the general symmetry principles.

In the following, we consider $C_\MR{3v}$ and $C_{4v}$, which are the point groups of the magnetic trimers and tetramers for which we will present results for the magnetic interactions.
The corresponding magnetic structures and symmetry operations are illustrated in Fig.~\ref{fig:symmetries_illustration} and the latter listed in Appendix~\ref{app:symm}.

\begin{figure}[tb]
\begin{center}
	\includegraphics[width=\columnwidth]{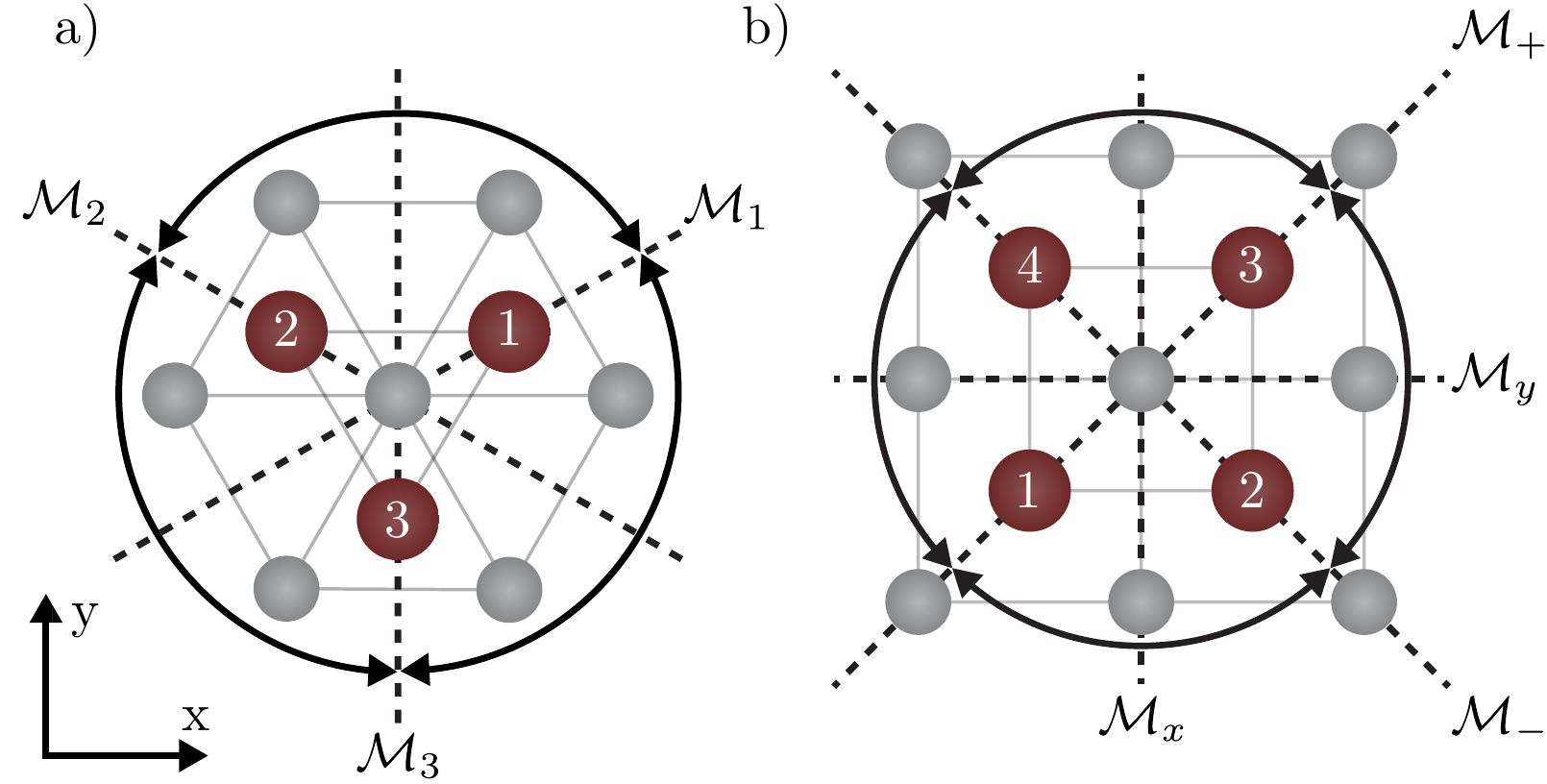}
	\caption{
	Illustration of the symmetries of the considered nanostructures.
	The magnetic atoms (numbered) are illustrated by red spheres and the surface atoms by grey spheres.
	Mirror symmetries are indicated by dashed lines while the arrows indicate rotational symmetries.
	a) Compact trimer on a hexagonal surface with $C_\MR{3v}$ symmetry.
	b) Compact tetramer on a square lattice with $C_\MR{4v}$ symmetry.
	}\label{fig:symmetries_illustration}
\end{center}
\end{figure}

\subsection{Symmetries for a trimer}\label{sec:c3v}
For the trimer ($C_\MR{3v}$ symmetry, see Fig.~\ref{fig:symmetries_illustration}a), the three mirror symmetries correspond to
\begin{align}
    \label{eq:c3vm1}
    (\VEC{S}_1,\VEC{S}_2,\VEC{S}_3) &\rightarrow \MC{M}_1(\VEC{S}_1,\VEC{S}_3,\VEC{S}_2) \; , \\
    \label{eq:c3vm2}
    (\VEC{S}_1,\VEC{S}_2,\VEC{S}_3) &\rightarrow \MC{M}_2(\VEC{S}_3,\VEC{S}_2,\VEC{S}_1) \; , \\
    \label{eq:c3vm3}
    (\VEC{S}_1,\VEC{S}_2,\VEC{S}_3) &\rightarrow \MC{M}_3(\VEC{S}_2,\VEC{S}_1,\VEC{S}_3) \; .
\end{align}
The isotropic interactions are only affected by a permutation of site labels, which implies $J_{12} = J_{13} = J_{23}$ and likewise for the isotropic biquadratic interaction $B_{ij}$.
Similarly, there is only one independent parameter for the isotropic 3-site interactions, $B_{123}$.

To illustrate a symmetry operation that maps a set of sites onto itself, consider the DMI between atoms 1 and 2 and the mirror symmetry $\MC{M}_3$.
The cross product transforms as $\VEC{S}_1 \times \VEC{S}_2 \rightarrow (\MC{M}_3 \VEC{S}_2) \times (\MC{M}_3 \VEC{S}_1) = \MC{M}_3 \left(\VEC{S}_1 \times \VEC{S}_2\right)$.
The DMI thus transforms as
\begin{equation}
    \VEC{D}_{12} \cdot \left(\VEC{S}_1 \times \VEC{S}_2\right) \rightarrow \left(\MC{M}_3\VEC{D}_{12}\right) \cdot \left(\VEC{S}_1 \times \VEC{S}_2\right) \;,
\end{equation}
which implies that $\MC{M}_3\VEC{D}_{12} = \VEC{D}_{12}$ for the energy to remain invariant, or $\VEC{D}_{12} = \left(0,D^y_{12},D^z_{12}\right)$.
This shows that $\VEC{D}_{12}$ lies in the mirror plane $\MC{M}_3$, as expected from Moriya's rules.
The mirror symmetries $\MC{M}_1$ and $\MC{M}_2$ map the atom pair $(1,2)$ into $(1,3)$ and into $(3,2)$, respectively, so they exemplify symmetries that relate different sites.
From these symmetries we get
\begin{align}
    \VEC{D}_{12} \cdot \left(\VEC{S}_1 \times \VEC{S}_2\right) &\rightarrow \left(\MC{M}_1\VEC{D}_{12}\right) \cdot \left(\VEC{S}_3 \times \VEC{S}_1\right) \;, \\
    \VEC{D}_{12} \cdot \left(\VEC{S}_1 \times \VEC{S}_2\right) &\rightarrow \left(\MC{M}_2\VEC{D}_{12}\right) \cdot \left(\VEC{S}_2 \times \VEC{S}_3\right) \;,
\end{align}
or $\MC{M}_1\VEC{D}_{12} = \VEC{D}_{31}$ and $\MC{M}_2\VEC{D}_{12} = \VEC{D}_{23}$.
The rotations lead to $\VEC{D}_{23} = \MC{R}_+\VEC{D}_{12}$ and $\VEC{D}_{31} = \MC{R}_-\VEC{D}_{12}$.
As we demonstrated in Ref.~\onlinecite{Brinker2019}, the CBI vector (see Eq.~\eqref{eq:2site}) has the same transformation properties as the DMI vector.
We thus have $\VEC{C}_{12} = \left(0,C^y_{12},C^z_{12}\right)$, and all the other vectors can be generated from this one with the same symmetry operations used for the DMI vectors.
If the symmetry is increased from $C_\MR{3v}$ to $D_\MR{3h}$ then the mirror symmetry $\MC{M}_z$ ($z \rightarrow -z$) also applies, which would lead to $-\MC{M}_z\VEC{D}_{12} = \VEC{D}_{12}$ and so $\VEC{D}_{12} = \left(0,0,D^z_{12}\right)$, again in accordance with Moriya's rules.

Similar considerations allow us to establish the symmetry properties of the chiral 3-site interaction vectors $\VEC{C}_{ijk}$ (see Eq.~\eqref{eq:3site}).
Take the interaction connecting atoms 1 and 3 through atom 2 as an example (see Fig.~\ref{fig:symmetries_illustration}a).
If SOC mediates the interaction between atoms 1 and 2 the interactions corresponds to Eq.~\eqref{eq:chiral_3-site_diagram} with the coefficient $\VEC{C}_{123}$.
In contrast to the isotropic multi-site interactions, as well as the DMI, there is no general relation between $\VEC{C}_{123}$ and $\VEC{C}_{321}$, as justified by the microscopic model that assigns SOC to a specific bond.
In addition, the $C_\MR{3v}$ point group symmetry does not map the form $\VEC{S}_1\times \VEC{S}_2 \left(\VEC{S}_2\cdot\VEC{S}_3\right)$ onto itself, which would be necessary in order to find symmetry constraints for $\VEC{C}_{123}$.
Thus, the interaction vector $\VEC{C}_{123}$ for the trimer is a general 3-component vector.
To show how a constraint on this interaction can emerge, consider increasing hypothetically the symmetry to $D_\MR{3h}$.
The additional mirror symmetry $\MC{M}_z$ leaves the site labels invariant and enforces $\VEC{C}_{123} = - \MC{M}_z \VEC{C}_{123}$, or $\VEC{C}_{123} = (0,0,C_{123}^z)$.

\begin{figure}[tb]
    \centering
    \includegraphics[width=0.7\columnwidth]{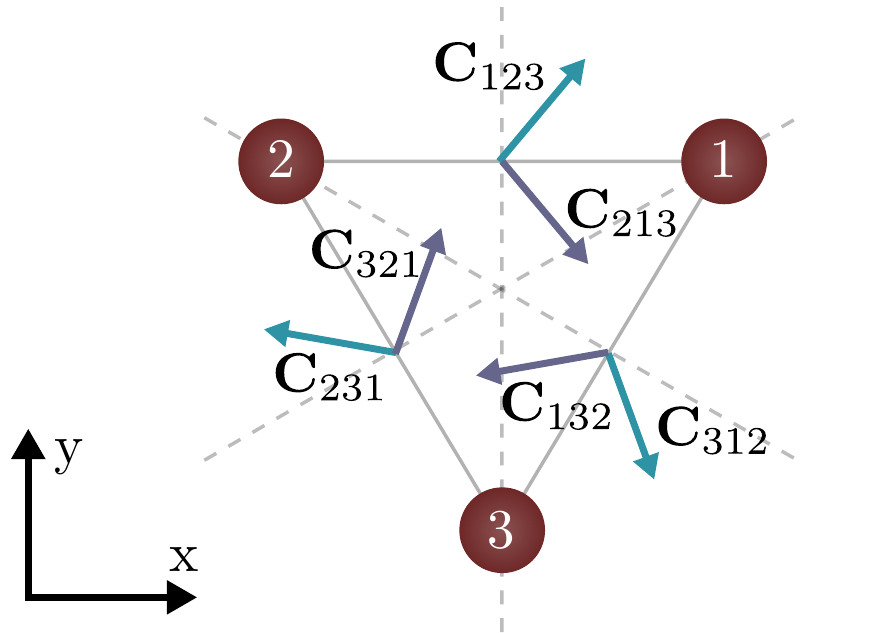}
    \caption{Relations between the chiral 3-site interaction vectors under $C_\MR{3v}$ symmetry.
    For comparison, the DMI vectors are along the mirror planes intersecting each edge of the triangle.}
    \label{fig:trianglechiral}
\end{figure}

Nonetheless, the $C_\MR{3v}$ symmetry of the trimer leads to a simplification of the chiral 3-site interaction since it can be used to relate the different interaction vectors to each other.
Starting from Eq.~\eqref{eq:3site} we can group the interactions in two subsets corresponding to a cyclic permutation of the sites, $\MC{S}_1 = \{ \VEC{C}_{123}, \VEC{C}_{231}, \VEC{C}_{312} \}$ and $\MC{S}_2 = \{ \VEC{C}_{321},\VEC{C}_{132},\VEC{C}_{213} \}$.
The rotational symmetries can be used to relate the vectors in the set $\MC{S}_1$ to each other, $\VEC{C}_{231} = \MC{R}_+ \VEC{C}_{123}$ and $\VEC{C}_{312} = \MC{R}_- \VEC{C}_{123}$.
The vectors in the set $\MC{S}_2$ transform among themselves in the same way, and the mirror symmetries connect the two sets.
For instance, the mirror $\MC{M}_3$ imposes
\begin{align}
    &\VEC{C}_{123} \cdot \left(\VEC{S}_1\times \VEC{S}_2\right) \left(\VEC{S}_2\cdot\VEC{S}_3\right) \nonumber\\
    &\rightarrow
    -\left(\MC{M}_3\VEC{C}_{123}\right) \cdot \left(\VEC{S}_2\times \VEC{S}_1\right) \left(\VEC{S}_1\cdot\VEC{S}_3\right) \quad ,
\end{align}
which leads to $\VEC{C}_{213} = -\MC{M}_3\VEC{C}_{123}$, and similarly $\VEC{C}_{321} = -\MC{M}_1\VEC{C}_{231}$ and $\VEC{C}_{132} = -\MC{M}_2\VEC{C}_{312}$.
The pattern formed by the six chiral vectors is shown in Fig.~\ref{fig:trianglechiral}.

We now illustrate the convention of Eq.~\eqref{eq:magen} for the 3-site interactions with the case of the trimer, for which summing over all triples results in
\begin{widetext}
\begin{align}\label{eq:energy3trimer}
    \MC{E}^{(3)} = \frac{1}{2}\;{\!\sum_{i,j,k}\!}' \MC{E}_{ijk} 
    &= B_{123}\,\big((\VEC{S}_1\cdot\VEC{S}_2) (\VEC{S}_2\cdot\VEC{S}_3) + (\VEC{S}_2\cdot\VEC{S}_3) (\VEC{S}_3\cdot\VEC{S}_1)
    + (\VEC{S}_3\cdot\VEC{S}_1) (\VEC{S}_1\cdot\VEC{S}_2)\big) \nonumber\\[-1.25em]
    &+ \frac{1}{2}\,\big(\VEC{C}_{123} \cdot (\VEC{S}_1\times\VEC{S}_2) (\VEC{S}_2\cdot\VEC{S}_3)
    + \VEC{C}_{231} \cdot (\VEC{S}_2\times\VEC{S}_3) (\VEC{S}_3\cdot\VEC{S}_1)
    + \VEC{C}_{312} \cdot (\VEC{S}_3\times\VEC{S}_1) (\VEC{S}_1\cdot\VEC{S}_2) \nonumber\\
    &\hspace{1.4em} + \VEC{C}_{321} \cdot (\VEC{S}_3\times\VEC{S}_2) (\VEC{S}_2\cdot\VEC{S}_1)
    + \VEC{C}_{132} \cdot (\VEC{S}_1\times\VEC{S}_3) (\VEC{S}_3\cdot\VEC{S}_2) 
    + \VEC{C}_{213} \cdot (\VEC{S}_2\times\VEC{S}_1) (\VEC{S}_1\cdot\VEC{S}_3)\big) \; .
\end{align}
\end{widetext}
We repeat that all chiral interaction vectors can be related to the general vector $\VEC{C}_{123}$ using the $C_\MR{3v}$ symmetry.

\subsection{Symmetries for a tetramer}\label{sec:c4v}
For the tetramer ($C_\MR{4v}$ symmetry, see Fig.~\ref{fig:symmetries_illustration}b), the four mirror symmetries correspond to the following mappings:
\begin{align}
    \label{eq:c4vmx}
    (\VEC{S}_1,\VEC{S}_2,\VEC{S}_3,\VEC{S}_4) &\rightarrow \MC{M}_x(\VEC{S}_2,\VEC{S}_1,\VEC{S}_4,\VEC{S}_3) \; , \\
    \label{eq:c4vmy}
    (\VEC{S}_1,\VEC{S}_2,\VEC{S}_3,\VEC{S}_4) &\rightarrow \MC{M}_y(\VEC{S}_4,\VEC{S}_3,\VEC{S}_2,\VEC{S}_1) \; , \\
    \label{eq:c4vmp}
    (\VEC{S}_1,\VEC{S}_2,\VEC{S}_3,\VEC{S}_4) &\rightarrow \MC{M}_+(\VEC{S}_1,\VEC{S}_4,\VEC{S}_3,\VEC{S}_2) \; , \\
    \label{eq:c4vmm}
    (\VEC{S}_1,\VEC{S}_2,\VEC{S}_3,\VEC{S}_4) &\rightarrow \MC{M}_-(\VEC{S}_3,\VEC{S}_2,\VEC{S}_1,\VEC{S}_4) \; .
\end{align}

There are two types of interactions, those that connect atoms only along the edges of the tetramer (nearest-neighbors) and those that include connections across the diagonals (next-nearest-neighbors).
For the 2-site interactions connecting $(i,j)$ we can list 8 ordered pairs along the edges and 4 ordered pairs along the diagonals, with $(j,i)$ being related to $(i,j)$ by construction (e.g.\ $J_{ji} = J_{ij}$ or $\VEC{D}_{ji} = -\VEC{D}_{ij}$).
We have $J_{12} = J_{23} = J_{34} = J_{41}$ (nearest-neighbors) and $J_{13} = J_{24}$ (next-nearest-neighbors), and similarly for the isotropic biquadratic interaction $B_{ij}$.
The DMI vectors along the edges of the tetramer have the same form as for the trimer, as they contain a mirror plane perpendicular to the edge connecting each pair, $\MC{M}_x\VEC{D}_{12} = \VEC{D}_{12}$ so $\VEC{D}_{12} = \left(0,D^y_{12},D^z_{12}\right)$.
The pattern of DMI vectors around the edges of the tetramer can then be simply obtained by rotation starting from $\VEC{D}_{12}$ as reference: $\VEC{D}_{23} = \MC{R}\VEC{D}_{12}$, $\VEC{D}_{34} = \MC{R}\VEC{D}_{23}$ and $\VEC{D}_{41} = \MC{R}\VEC{D}_{34}$.
The DMI vector across the diagonals have to comply with two mirror planes, $\MC{M}_+$ and $\MC{M}_-$.
We have $\VEC{D}_{24} = -\MC{M}_-\VEC{D}_{24}$ and $\VEC{D}_{24} = +\MC{M}_+\VEC{D}_{24}$ so $\VEC{D}_{24} = D_{24} \left(\frac{1}{\sqrt{2}},\frac{1}{\sqrt{2}},0\right)$, and the other DMI vector is obtained from $\VEC{D}_{31} = \MC{M}_x\VEC{D}_{24} = D_{24} \left(-\frac{1}{\sqrt{2}},\frac{1}{\sqrt{2}},0\right)$.
Once again, the CBI vector $\VEC{C}_{ij}$ has the same properties as the DMI vector.

The 3-site interactions are specified by triples $(i,j,k)$, for which 8 connect only sites along the edges (nearest-neighbors) and 16 include a diagonal connection (next-nearest-neighbors).
For the isotropic 3-site interactions given in Eq.~\eqref{eq:3site} we have $B_{ijk} = B_{kji}$ (the repeated site is unchanged) and two independent parameters: $B_{123}$ if both dot products are along the edges, and $B_{124}$ if one of the dot products is along the diagonal.

The chiral 3-site interactions given in Eq.~\eqref{eq:3site} can be separated into three groups.
The diagrams from our heuristic arguments provide visual insight into this:
\begin{align}
    \label{eq:43g1}
    \begin{tikzpicture}[baseline={([yshift=-.5ex]current bounding box.center)}]
      \begin{scope}[scale=2]
          % coordinates
          \coordinate (lb) at (0,0);
          \coordinate (rb) at (0.5,0);
          \coordinate (rt) at (0.5,0.5);
          \coordinate (lt) at (0.0,0.5);
          % edges
          \draw (lb) edge[Green_soc,out=-30,in=-150] node[left,xshift=-.2cm] {} (rb);
          \draw (rb) edge[Green,out=60,in=-60] node[right,xshift=.2cm] {} (rt);
          \draw (rt) edge[Green,out=-120,in=120] node[right,yshift=.2cm] {} (rb);
          \draw (rb) edge[Green,out=150,in=30] node[left,yshift=.2cm] {} (lb);
          % arrows
          \draw[susc] (lb) circle (1pt) node[left,yshift=-.3cm] {$1$};
          \draw[susc] (rb) circle (1pt) node[right,yshift=-.3cm] {$2$};
          \draw[susc] (rt) circle (1pt) node[right,yshift=.3cm] {$3$};
      \end{scope}
    \end{tikzpicture}
    &\rightarrow \quad \VEC{C}_{123} \cdot \left(\VEC{S}_1\times \VEC{S}_2\right) \left(\VEC{S}_2\cdot\VEC{S}_3\right)
    \; , \\
    \label{eq:43g2}
    \begin{tikzpicture}[baseline={([yshift=-.5ex]current bounding box.center)}]
      \begin{scope}[scale=2]
          % coordinates
          \coordinate (lb) at (0,0);
          \coordinate (rb) at (0.5,0);
          \coordinate (rt) at (0.5,0.5);
          \coordinate (lt) at (0.0,0.5);
          % edges
          \draw (lb) edge[Green_soc,out=-30,in=-150] node[left,xshift=-.2cm] {} (rb);
          \draw (rb) edge[Green,out=110,in=-20] node[right,xshift=.2cm] {} (lt);
          \draw (lt) edge[Green,out=-70,in=160] node[right,yshift=.2cm] {} (rb);
          \draw (rb) edge[Green,out=150,in=30] node[left,yshift=.2cm] {} (lb);
          % arrows
          \draw[susc] (lb) circle (1pt) node[left,yshift=-.3cm] {$1$};
          \draw[susc] (rb) circle (1pt) node[right,yshift=-.3cm] {$2$};
          \draw[susc] (lt) circle (1pt) node[left,yshift=.3cm] {$4$};
      \end{scope}
    \end{tikzpicture}
    &\rightarrow \quad \VEC{C}_{124} \cdot \left(\VEC{S}_1\times \VEC{S}_2\right) \left(\VEC{S}_2\cdot\VEC{S}_4\right)
    \; , \\
    \label{eq:43g3}
    \begin{tikzpicture}[baseline={([yshift=-.5ex]current bounding box.center)}]
      \begin{scope}[scale=2]
          % coordinates
          \coordinate (lb) at (0,0);
          \coordinate (rb) at (0.5,0);
          \coordinate (rt) at (0.5,0.5);
          \coordinate (lt) at (0.0,0.5);
          % edges
          \draw (lb) edge[Green,out=-30,in=-150] node[left,xshift=-.2cm] {} (rb);
          \draw (rb) edge[Green,out=110,in=-20] node[right,xshift=.2cm] {} (lt);
          \draw (lt) edge[Green_soc,out=-70,in=160] node[right,yshift=.2cm] {} (rb);
          \draw (rb) edge[Green,out=150,in=30] node[left,yshift=.2cm] {} (lb);
          % arrows
          \draw[susc] (lb) circle (1pt) node[left,yshift=-.3cm] {$1$};
          \draw[susc] (rb) circle (1pt) node[right,yshift=-.3cm] {$2$};
          \draw[susc] (lt) circle (1pt) node[left,yshift=.3cm] {$4$};
      \end{scope}
    \end{tikzpicture}
    &\rightarrow \quad \VEC{C}_{421} \cdot \left(\VEC{S}_4\times \VEC{S}_2\right) \left(\VEC{S}_2\cdot\VEC{S}_1\right)
    \; .
\end{align}
The dashed line indicates the location of the cross product between spins arising from SOC.
The $C_\MR{4v}$ symmetry places no constraints on a single chiral interaction vector $\VEC{C}_{ijk}$, but establishes groups of vectors which are related to each other by symmetry.

The diagram given in Eq.~\eqref{eq:43g1} has connections with both cross and dot products along the edges (nearest-neighbors).
It can be drawn in 8 symmetry-related ways, and the corresponding interaction vectors are all related to $\VEC{C}_{123}$.
The rotations give directly $\VEC{C}_{234} = \MC{R} \VEC{C}_{123}$, $\VEC{C}_{341} = \MC{R} \VEC{C}_{234}$, and $\VEC{C}_{412} = \MC{R} \VEC{C}_{341}$.
The remaining four chiral vectors can be related to $\VEC{C}_{123}$ using a mirror, $\VEC{C}_{214} = -\MC{M}_x \VEC{C}_{123}$, and rotations, $\VEC{C}_{321} = \MC{R} \VEC{C}_{214}$, $\VEC{C}_{432} = \MC{R} \VEC{C}_{321}$ and $\VEC{C}_{143} = \MC{R} \VEC{C}_{432}$.
The connections including a diagonal have to be distinguished by whether the cross product occurs on an edge or on a diagonal.
The first case is represented by the diagram in Eq.~\eqref{eq:43g2}, which can be drawn in 8 symmetry-related ways, with all chiral vectors being related to $\VEC{C}_{124}$.
The rotations give $\VEC{C}_{231} = \MC{R} \VEC{C}_{124}$, $\VEC{C}_{342} = \MC{R} \VEC{C}_{231}$, and $\VEC{C}_{413} = \MC{R} \VEC{C}_{342}$.
Applying a mirror we find $\VEC{C}_{213} = -\MC{M}_x \VEC{C}_{124}$, and with rotations we get $\VEC{C}_{324} = \MC{R} \VEC{C}_{213}$, $\VEC{C}_{431} = \MC{R} \VEC{C}_{324}$ and $\VEC{C}_{142} = \MC{R} \VEC{C}_{431}$.
The second case is represented by the diagram in Eq.~\eqref{eq:43g3}, which can be drawn in 8 symmetry-related ways, with all chiral vectors being related to $\VEC{C}_{421}$.
The rotations give $\VEC{C}_{132} = \MC{R} \VEC{C}_{421}$, $\VEC{C}_{243} = \MC{R} \VEC{C}_{132}$, and $\VEC{C}_{314} = \MC{R} \VEC{C}_{243}$.
Applying a mirror we find $\VEC{C}_{423} = -\MC{M}_- \VEC{C}_{421}$, and with rotations we get $\VEC{C}_{134} = \MC{R} \VEC{C}_{423}$, $\VEC{C}_{241} = \MC{R} \VEC{C}_{134}$ and $\VEC{C}_{312} = \MC{R} \VEC{C}_{241}$.
This completes the list of the chiral 3-site vectors.

The 4-site interactions are specified by quadruples $(i,j,k,l)$, for which 8 connect only sites along the edges (nearest-neighbors) and 16 include a diagonal connection (next-nearest-neighbors).
For the isotropic 4-site interactions given in Eq.~\eqref{eq:3site} we have $B_{ijkl} = B_{jikl} = B_{ijlk} = B_{jilk}$ and two independent parameters: $B_{1234}$ if both dot products are along the edges, and $B_{1324}$ if both dot products are along the diagonal.

For the chiral 4-site interactions given in Eq.~\eqref{eq:3site} we have $\VEC{C}_{ijkl} = \VEC{C}_{ijlk} = -\VEC{C}_{jikl} = -\VEC{C}_{jilk}$, which reduces the amount of independent terms by a factor of four.
The six independent terms can be further grouped according to whether the cross product is along an edge of the tetramer (four terms) or along a diagonal (two terms), each generated by a reference interaction vector.
In contrast to the chiral 3-site interactions, the reference chiral vectors have additional constraints enforced by the mirror symmetries.
The chiral interaction vectors for the group of four terms are generated from $\VEC{C}_{1234}$ by rotational symmetry: $\VEC{C}_{2341} = \MC{R}\VEC{C}_{1234}$, $\VEC{C}_{3412} = \MC{R}\VEC{C}_{2341}$ and $\VEC{C}_{4123} = \MC{R}\VEC{C}_{3412}$.
The reference chiral vector is constrained by the mirror plane $\MC{M}_x$ combined with the general symmetry of the interaction, which yields $-\MC{M}_x \VEC{C}_{1234} = \VEC{C}_{2143} = -\VEC{C}_{1234}$, and so $\VEC{C}_{1234} = (0, C_{1234}^y,C_{1234}^z)$.
This is analogous to the Moriya rule for the DMI vector: if the bond containing the cross product is bisected by a mirror plane, the chiral vector must lie in this mirror plane.
The chiral interaction vectors for the group of two terms are specified by $\VEC{C}_{1324}$ and $\VEC{C}_{2431} = \MC{R}\VEC{C}_{1324}$.
Now the $\MC{M}_+$ symmetry imposes a stronger constraint on the reference vector, $ -\MC{M}_+ \VEC{C}_{1324} = \VEC{C}_{1342} = \VEC{C}_{1324}$, from which follows $\VEC{C}_{1324} = C_{1324} \left(-\frac{1}{\sqrt{2}}, \frac{1}{\sqrt{2}}, 0\right)$.
This is analogous to the Moriya rule for the DMI vector: if the bond containing the cross product lies in a mirror plane, the chiral vector must be perpendicular to this mirror plane.
If we again consider increasing hypothetically the symmetry, this time to $D_\MR{4h}$, the additional mirror symmetry $\MC{M}_z$ leaves the site labels invariant and enforces $\VEC{C}_{1234} = - \MC{M}_z \VEC{C}_{1234}$, or $\VEC{C}_{1234} = (0,0,C_{1234}^z)$.
Thus, the chiral 4-site interaction vector $\VEC{C}_{ijkl}$ obeys symmetry rules similar to the well-known Moriya rules for the DMI vector, if the symmetry operation maps the pairs $(i,j)$ and $(k,l)$ onto themselves (up to reordering).

We now illustrate the convention of Eq.~\eqref{eq:magen} for the 4-site interactions with the case of the tetramer, for which summing over all quadruples results in
\begin{widetext}
\begin{align}\label{eq:energy4tetramer}
    \MC{E}^{(4)} = \frac{1}{4}\,{\!\sum_{i,j,k,l}\!}' \MC{E}_{ijkl} 
    &= 2B_{1234}\,\big((\VEC{S}_1\cdot\VEC{S}_2) (\VEC{S}_3\cdot\VEC{S}_4) + (\VEC{S}_1\cdot\VEC{S}_4) (\VEC{S}_2\cdot\VEC{S}_3)\big)
    + 2B_{1324}\,(\VEC{S}_1\cdot\VEC{S}_3) (\VEC{S}_2\cdot\VEC{S}_4) \nonumber\\[-1.25em]
    &+ \VEC{C}_{1234} \cdot (\VEC{S}_1\times\VEC{S}_2) (\VEC{S}_3\cdot\VEC{S}_4)
    + \VEC{C}_{2341} \cdot (\VEC{S}_2\times\VEC{S}_3) (\VEC{S}_4\cdot\VEC{S}_1)
    + \VEC{C}_{3412} \cdot (\VEC{S}_3\times\VEC{S}_4) (\VEC{S}_1\cdot\VEC{S}_2) \nonumber\\
    &+ \VEC{C}_{4123} \cdot (\VEC{S}_4\times\VEC{S}_1) (\VEC{S}_2\cdot\VEC{S}_3) 
    + \VEC{C}_{1324} \cdot (\VEC{S}_1\times\VEC{S}_3) (\VEC{S}_2\cdot\VEC{S}_4)
    + \VEC{C}_{2413} \cdot (\VEC{S}_2\times\VEC{S}_4) (\VEC{S}_1\cdot\VEC{S}_3) \; .
\end{align}
\end{widetext}

\section{Global mapping from DFT calculations to the spin model}\label{sec:mapping}
The energy of a chosen magnetic structure for a given material can be obtained from first principles, for instance from a DFT calculation.
Suppose that the relevant coarse-grained variables which are necessary to map to the spin model are already defined.
These are the localized spin moments $\VEC{M}_i$ that exist on some subset of all the atomic sites of the magnetic material.
The spin moments must be relatively rigid, so that their magnitude is not strongly dependent on the magnetic structure.
It is then meaningful to separate the magnitude from the orientation, $\VEC{M}_i = M_i\,\VEC{S}_i$ (recall $|\VEC{S}_i| = 1$), and to identify the orientations of the magnetic moments with those in the target classical spin model.
One can then view the DFT total energy as a functional of the magnetic structure specified by those orientations, $E^\MR{DFT}[\VEC{S}_1,\ldots,\VEC{S}_N]$.
In practice, an arbitrary magnetic structure is not a stationary solution of the DFT total energy, which must then be stabilized by the addition of constraints \cite{Dederichs1984,Ujfalussy1999,Kurz2004},
\begin{equation}
    E^\MR{cDFT} = E^\MR{DFT} + \sum_i \VEC{S}_i \cdot \VEC{B}_i
\end{equation}
satisfying the condition
\begin{equation}
    \frac{\partial E^\MR{cDFT}}{\partial\VEC{S}_i} = \frac{\partial E^\MR{DFT}}{{\partial\VEC{S}_i}} + \VEC{B}_i = 0
\end{equation}
when the derivatives are evaluated for the chosen magnetic structure $\{\VEC{S}_1,\ldots,\VEC{S}_N\}$.
The self-consistently obtained constraining magnetic field $\VEC{B}_i$ is thus equal to the exact total energy derivative with respect to the orientation of the spin moment on site $i$, and this is the key observation that is used to map the DFT calculations to the classical spin model given in Eq.~\eqref{eq:magen},
\begin{equation}
    \VEC{B}_i = -\frac{\partial E^\MR{DFT}}{{\partial\VEC{S}_i}} \underset{\MR{mapping}}{=} -\frac{\partial \MC{E}}{{\partial\VEC{S}_i}} \quad .
\end{equation}

The dependence of the DFT total energy on the spin orientations can be arbitrarily complex, and so the mapping must be chosen in a way that allows for systematic improvement and a controllable error between the energy obtained from DFT and the one computed from the parametrized spin model, for a selected set of magnetic structures.
We have to generate a number of magnetic structures that is enough to obtain all the magnetic interaction coefficients from fitting the self-consistent constraining magnetic fields from the respective DFT calculations to the corresponding derivatives of the spin model.
To this end, each spin moment is set to one of 14 predefined orientations (6 along the cartesian axes plus 8 along the diagonals of each octant).
The magnetic structures are constructed by the tensor product of all possible orientations of each spin moment, leading to $14^N$ magnetic structures.
For the systems we consider, we would have 2744 magnetic structures for the trimer ($N = 3$) and 38416 magnetic structures for the tetramer ($N = 4$).
Enforcing time-reversal symmetry together with the corresponding point group symmetry ($C_\MR{3v}$ for the trimer and $C_\MR{4v}$ for the tetramer) achieves a reduction to 252 and 2513 inequivalent magnetic structures, respectively.
As the number of configurations is still fairly large for the tetramer, we resort to randomly sampling 250 magnetic structures from this subset of inequivalent magnetic structures.
The constraining magnetic fields for each magnetic structure are then self-consistently obtained from the corresponding DFT calculation.

The quality of the fit is quantified by the mean-average-error (mae) between the constraining fields and the corresponding derivatives of the fitted spin model, summed over the used magnetic structures:
\begin{equation} \label{eq:mae}
    \MR{mae} = \frac{1}{N_s N_a} \sum_{s=1}^{N_s} \sum_{i=1}^{N_a} \left|\frac{\partial E^\MR{DFT}}{{\partial\VEC{S}_i}} - \frac{\partial \MC{E}}{{\partial\VEC{S}_i}}\right|_{\{\VEC{S}\}_s} \; .
\end{equation}
Here the sum is over the used $N_s$ magnetic structures, and for each magnetic structure $\{\VEC{S}\}_s$ we compute for all the $N_a$ atoms in the trimer or tetramer the absolute error between cDFT and the fitted spin model.

\section{Computational Details}
        We employ the all-electron Korringa-Kohn-Rostoker Green function method in full potential with spin-orbit coupling added to the scalar relativistic approximation \cite{Papanikolaou2002,Bauer2014}.
        Exchange and correlation effects are treated in the local spin density approximation (LSDA) as parametrized by Vosko, Wilk and Nusair \cite{Vosko1980}.
        The pure surfaces are modeled by 22 layers with two vacuum regions corresponding to four inter-layer distances each using the experimental lattice constants, which are for the considered fcc structures $a^\mathrm{Pt} = \SI{3.924}{\angstrom}$ and $a^\mathrm{Au} = \SI{4.078}{\angstrom}$, and for the considered hcp structure $a^\mathrm{Re} = \SI{2.761}{\angstrom}$ and $c^\mathrm{Re} = \SI{4.456}{\angstrom}$.
        The scattering wave functions are expanded up to an angular momentum cutoff of $\ell_\text{max}=3$ and a $k$-mesh of $150\times150$ is used.
        The nanostructures are embedded in real space using a nearest-neighbor cluster.
        The constrained DFT calculations for the magnetic structures required for the mapping to the spin model are performed as explained in the previous section, following the procedure introduced for magnetic dimers in our previous work, Ref.~\onlinecite{Brinker2019}.
        
        To account for structural relaxations we use the Quantum Espresso package \cite{QE2009,QE2017}.
        The surfaces are modelled by five layers surrounded by a vacuum region of the same thickness.
        The nanostructures are placed on $4\times4$ supercells of the surfaces and a $k$-mesh of $2\times2\times1$ is used.
        Exchange and correlations effects are treated in the generalized gradient approximation using the PBEsol functional \cite{Perdew2008}, and we used ultrasoft pseudopotentials from the pslibrary\cite{pslibrary} with an energy cutoff of \SI{100}{\rydberg}.
        The nanostructure as well as the first surface layer are allowed to relax.
        For the setup of the KKR geometry we use only the vertical relaxation of the nanostructure, since the relaxation of the first surface layer turns out to be negligible in agreement with previous studies \cite{Blonski2009}.
        The results of the structural relaxations are summarized in Table~\ref{tab:relaxations}.

        \begin{table}[tb]
        \centering
        \begin{ruledtabular}
            \begin{tabular}{clrrrr}
            Surface & System & $r_\mathrm{QE}$ [\%]  & $M_\mathrm{QE} [\mu_\text{B}]$ & $r_\mathrm{KKR}$ [\%] & $M_\mathrm{KKR} [\mu_\text{B}]$\\ \hline
                \multirow{4}{*}{Pt(111)} & Cr$_3$ & 17.3 & 3.09 &  \multirow{4}{*}{17.5} & 3.33\\%\cline{2-4}
                 & Mn$_3$ & 15.9 & 3.76 & & 4.05\\%\cline{2-4}
                 & Fe$_3$ & 18.1 & 3.08 & & 3.24 \\%\cline{2-4}
                 & Co$_3$ & 19.4 & 2.11 & & 2.12 \\\hline
                \multirow{2}{*}{Pt(001)} & Cr$_4$ & 21.6 & 2.93 &   \multirow{2}{*}{22.5} & 3.04 \\%\cline{2-4}
                 & Fe$_4$ & 24.4 & 3.25 &  & 3.19  \\\hline
                 \multirow{4}{*}{Re(0001)} & Cr$_3$ & 20.3 & 1.50 & \multirow{4}{*}{15.0} & 1.80 \\%\cline{2-4}
                 & Mn$_3$ & 14.1 & 3.17 & & 3.07 \\
                 & Fe$_3$ & 15.5 & 2.58 & & 2.55 \\
                 & Co$_3$ & 16.7 & 1.31 & & 1.38 \\\hline
                \multirow{4}{*}{Au(111)} & Cr$_3$ & 15.4 & 3.61 & \multirow{4}{*}{17.5} & 4.09 \\%\cline{2-4}
                 & Mn$_3$ & 19.0 & 3.88 & & 4.26 \\%\cline{2-4}
                 & Fe$_3$ & 19.5 & 2.92 & & 3.24 \\%\cline{2-4}
                 & Co$_3$ & 19.5 & 2.00 & & 2.05 \\
            \end{tabular}
        \end{ruledtabular}
    	\caption{\label{tab:relaxations}
    	Magnetic ground state properties of different nanostructures obtained from Quantum Espresso and from KKR.
    	We considered compact fcc-top-stacked trimers for the Pt(111) and the Au(111) surfaces, and hcp-top-stacked trimers for the Re(0001) surface.
    	On the Pt(001) surface two compact tetramers are used.
    	We define the average relaxation of a nanostructure as $r = 1 - d/d_0$, where $d$ is the average vertical distance between the atoms comprising the nanostructure and the atoms of the surface layer, and $d_0$ is the bulk vertical interlayer distance.
    	These are computed with Quantum Espresso, and inform the value used in the KKR calculations as shown.
    	We also show the spin magnetic moments per atom $M$ obtained from both types of calculations.
    	}
    	\label{tab:Geometries}
    \end{table}

\section{Results}\label{sec:results}
We consider Cr, Mn, Fe and Co trimers on the Pt(111), Re(0001) and Au(111) surfaces, and additionally Cr and Fe tetramers on the Pt(001) surface.
The constrained magnetic configurations described in Section~\ref{sec:mapping} were used to fit the interaction parameters for the 1-site, 2-site, 3-site and 4-site interactions.
The quality of the fits is quantified by the mean-average-error, Eq.~\eqref{eq:mae}.
We show in Table~\ref{tab:mae_trimer} how the fit improves (or not) by adding more types of interactions to the spin model.
A significant drop in the mean-average-error when adding a new class of interactions to the fit shows that these make an important contribution to the energy.
However, no significant improvement in the fitting error can also arise due to the weakness of the additionally-fitted interactions.
Overall, all of the systems could be well-fitted with our procedure.

\begin{table}[tb]
    \centering
    \begin{ruledtabular}
    \begin{tabular}{cccccc}
        \multicolumn{2}{r}{$(p\text{-spin},q\text{-site})$ interactions:} & $(2,2)$  & $(4,2)$ & $(4,3)$ &  $(4,4)$ \\ \hline
        \multirow{4}{*}{Pt(111)} & Cr$_3$ & 4.78  &    3.39  &    1.12  & ---  \\%\cline{2-4}
         & Mn$_3$ &  1.09  &    0.63  &    0.27  & --- \\%\cline{2-4}
         & Fe$_3$ &  2.35  &    2.16  &    0.80  & --- \\%\cline{2-4}
         & Co$_3$ &  1.29  &    1.18  &    0.91   & --- \\\hline
         \multirow{4}{*}{Re(0001)} & Cr$_3$ & 1.54  &    0.77  &    0.57  & --- \\%\cline{2-4}
         & Mn$_3$ &  2.06  &    1.63  &    1.19  & --- \\%\cline{2-4}
         & Fe$_3$ &  1.36  &    1.20  &    1.18  & --- \\%\cline{2-4}
         & Co$_3$ &  0.66  &    0.45  &    0.39  & --- \\\hline
        \multirow{4}{*}{Au(111)} & Cr$_3$ & 4.46  &    3.49  &    0.68   & --- \\%\cline{2-4}
         & Mn$_3$ &  1.38  &    0.43  &    0.38  & --- \\%\cline{2-4}
         & Fe$_3$ &  3.81  &    2.66  &    1.57  & --- \\%\cline{2-4}
         & Co$_3$ &  1.98  &    1.51  &    0.58  & --- \\\hline
        \multirow{2}{*}{Pt(001)} & Cr$_4$ & 6.60  &    2.49  &    1.95  &    1.69 \\%\cline{2-4}
         & Fe$_4$ &  2.85  &    1.62  &    1.41  &    1.26  \\%\cline{2-4}
    \end{tabular}
    \end{ruledtabular}
    \caption{\label{tab:mae_trimer}
    Mean-average-error (mae), Eq.~\eqref{eq:mae}, in units of [meV] of the spin model fits to the constrained DFT calculations for the different trimers and tetramers that we studied.
    }
\end{table}

\subsection{Trimers with $C_\MR{3v}$ symmetry}\label{sec:results_trimers}
A trimer of identical magnetic atoms arranged as a compact triangle on an fcc(111) or hcp(0001) surface has $C_\MR{3v}$ symmetry.
The chosen coordinate system, labelling of the magnetic atoms and mirror planes is shown in Fig.~\ref{fig:symmetries_illustration}a.
These are called top-stacked trimers, as they enclose a surface atom\cite{Hermenau2019}.
The isotropic and chiral interactions for the trimer are specified by Eqs.~\eqref{eq:2site} and \eqref{eq:3site}, with the detailed form for 3-site interactions given in Eq.~\eqref{eq:energy3trimer}.
The reference interaction parameters are given in Table~\ref{tab:trimer_exchange_parameters}, from which the full spin model can be parametrized by applying the $C_\MR{3v}$-symmetry operations.
The symmetric anisotropy parameters (see Eqs.~\eqref{eq:1site} and Eq.~\eqref{eq:2site} and Appendix~\ref{app:wizardry}) are not of our primary interest, but can be found in Table~\ref{tab:trimer_symmetric_aniso_parameters}.

    \begin{table*}[!t]
        \centering
        \begin{ruledtabular}
            \renewcommand{\arraystretch}{1.1}
            \begin{tabular}{lrrrrrrrrrrr}
            Surface & System & $J_{12}$ & $D_{12}^{y}$ & $D_{12}^{z}$ & $B_{12}$ & $C_{12}^{y}$ & $C_{12}^{z}$ & $B_{123}$ & $C_{123}^x$ & $C_{123}^y$ & $C_{123}^z$ \\ \hline
                \multirow{4}{*}{Pt(111)}   & Cr$_3$ &   74.35 &  -4.43 &  -4.49 &  -6.25 &  -1.85 &  -0.03 &   7.72 &  -0.54 &  -0.09 &  -2.23  \\%\cline{2-12}
                                           & Mn$_3$ &   62.58 &   5.68 &   0.78 &   1.72 &   0.97 &  -0.04 &   1.23 &   0.47 &  -0.09 &   0.23 \\%\cline{2-12}
                                           & Fe$_3$ &  -50.00 &   4.66 &   0.94 &  -1.21 &  -1.23 &   0.01 &   4.48 &   0.02 &   1.46 &  -1.31  \\%\cline{2-12}
                                           & Co$_3$ &  -66.24 &  -7.84 &   5.05 &   0.67 &   1.33 &  -0.15 &  -1.68 &  -0.22 &  -0.38 &   0.22  \\\hline
                 \multirow{4}{*}{Re(0001)} & Cr$_3$ &    9.33 &  -7.90 &  -0.78 &  -2.69 &   0.45 &  -0.23 &   0.55 &   1.05 &  -0.84 &   1.46  \\%\cline{2-12}
                                           & Mn$_3$ &   -3.72 &  -11.88 &   0.04 &  -2.63 &  -1.30 &   0.51 &   2.01 &  -1.55 &   1.41 &  -1.71 \\%\cline{2-12}
                                           & Fe$_3$ &   -16.14 &   1.52 &   2.12 &  -1.38 &   0.50 &  -0.53 &  -0.07 &   0.62 &  -0.52 &   0.40 \\%\cline{2-12}
                                           & Co$_3$ &    -5.66 &   4.47 &  -1.67 &  -0.98 &  -0.19 &   0.21 &  -0.43 &   0.06 &   0.29 &  -0.02 \\\hline
                \multirow{4}{*}{Au(111)}   & Cr$_3$ &   88.10 &   -3.73 &   1.69 &  -5.10 &  -1.07 &   0.53 &   8.06 &   0.22 &  -1.19 &  -0.06 \\%\cline{2-12}
                                           & Mn$_3$ &  -10.40 &   4.94 &   1.76 &   2.65 &   0.06 &  -0.27 &   0.25 &  -0.18 &   0.56 &  -0.25 \\%\cline{2-12}
                                           & Fe$_3$ & -113.04 &   3.66 &  -3.53 &  -5.52 &  -0.38 &   2.14 &   5.02 &   1.07 &  -1.06 &   1.71 \\%\cline{2-12}
                                           & Co$_3$ &  -72.41 &   0.34 &  -2.20 &   2.39 &   0.27 &  -0.80 &  -3.14 &  -0.78 &   1.12 &   0.79
            \end{tabular}
        \end{ruledtabular}
    	\caption{\label{tab:trimer_exchange_parameters}
    	Magnetic interactions in different compact top-stacked trimers in units of [meV].
    	We give the reference parameters for all isotropic and chiral 2-site and 3-site interactions, Eqs.~\eqref{eq:2site} and \eqref{eq:3site}.
    	The full set of interactions can be obtained from the shown ones by applying the $C_\MR{3v}$-symmetry operations to the trimer, see Section~\ref{sec:c3v}.
    	}
    \end{table*}

\begin{figure}[tb]
    \includegraphics[width=1.\columnwidth]{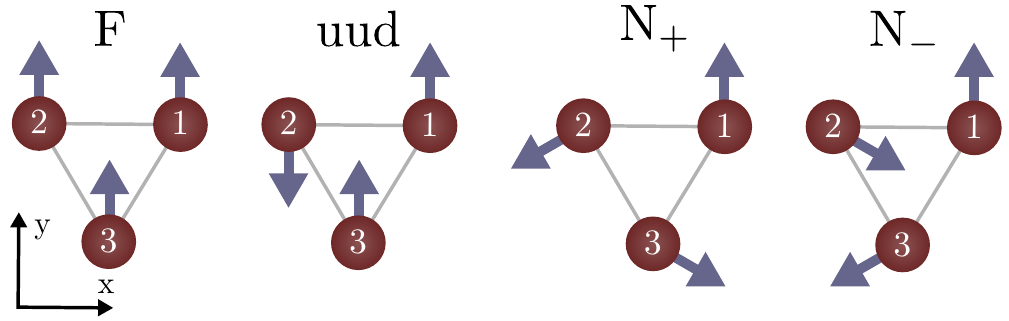}
    \caption{\label{fig:trimer_states}
    Basic magnetic structures for a trimer.
    F: Ferromagnetic state.
    uud: up-up-dow state.
    N$_+$: planar N\'eel state with anticlockwise rotation of the spins.
    N$_-$: planar N\'eel state with clockwise rotation of the spins.
    }
\end{figure}

We first discuss possible scenarios considering the isotropic interactions alone.
The corresponding magnetic structures are sketched in Fig.~\ref{fig:trimer_states}.
The dominant interaction is almost always the isotropic 2-spin interaction $J_{12}$.
$J_{12} < 0$ favors a ferromagnetic alignment, while $J_{12} > 0$ favors the planar N\'eel state for the trimer (the two forms $N_+$ and $N_-$ shown in Fig.~\ref{fig:trimer_states} have the same energy).
The isotropic biquadratic interactions $B_{12}$ and the 3-site ones $B_{123}$ tend to have comparable magnitudes, and mostly opposite signs.
$B_{12} < 0$ favors collinear states and $B_{12} > 0$ an $xyz$ state where all three spins are mutually perpendicular (not shown), while $B_{123} < 0$ favors a ferromagnetic alignment and $B_{123} > 0$ an up-up-down state.
Including all isotropic interactions, the ferromagnetic state has energy $\MC{E}_\MR{F} = 3\left(J_{12} + B_{12} + B_{123}\right)$, the $xyz$ state has zero energy, the up-up-down state $\MC{E}_\MR{uud} = -J_{12} + 3 B_{12} - B_{123}$, and the planar N\'eel states $\MC{E}_\MR{N} = \frac{3}{4}\left(-2 J_{12} + B_{12} + B_{123}\right)$.
The data in Table~\ref{tab:trimer_exchange_parameters} shows that for most trimers $B_{12} < 0$ and $B_{123} > 0$, so their combined action prefers an up-up-down state.
Its energy difference to the ferromagnetic state is $\MC{E}_\MR{uud} - \MC{E}_\MR{F} = -4\left(J_{12} + B_{123}\right)$.
We find this energy difference to be about $\SI{7}{\milli\electronvolt}$ for the Mn trimer on Re(0001), but for all other trimers with $J_{12} < 0$ the ferromagnetic state is much more stable than the other states.
If $J_{12} > 0$ then the relevant energy difference is to the planar N\'eel state, $\MC{E}_\MR{uud} - \MC{E}_\MR{N} = \frac{1}{4}\left(2 J_{12} + 9 B_{12} - 7 B_{123}\right)$, showing that the isotropic 4-spin interactions strongly penalize the planar N\'eel state.
For instance, this energy difference is just $\SI{10}{\milli\electronvolt}$ for Cr$_3$ on Pt(111), and becomes negative ($\SI{-2}{\milli\electronvolt}$) when this trimer is placed on Re(0001), which would make the up-up-down state the ground state if only isotropic interactions were at play.

In order to easily compare the isotropic and chiral contributions to the magnetic energy of the trimer, we now consider a family of magnetic structures with 3-fold rotational symmetry.
We parametrize the spins as
\begin{equation}\label{eq:magstruct}
    \VEC{S}_i = \cos\phi_i \sin\theta\,\hat{\VEC{x}} + \sin\phi_i \sin\theta\,\hat{\VEC{y}} + \cos\theta\,\hat{\VEC{z}} \;,
\end{equation}
with $\phi_1 = \phi + \SI{30}{\degree}$, $\phi_2 = \phi_1 + s\,\SI{120}{\degree}$ and $\phi_3 = \phi_1 + s\,\SI{240}{\degree}$.
For $\theta = \SI{0}{\degree}$ or \SI{180}{\degree} we have the ferromagnetic state along $\pm z$, and for $\theta = \SI{90}{\degree}$ we have the planar N\'eel state with either clockwise ($s = -$) or anticlockwise rotation of the spins ($s = +$), see Fig.~\ref{fig:trimer_states}.
We define the cosine $\alpha(\theta) = \VEC{S}_1\cdot\VEC{S}_2$ and sine $\beta(\theta) = |\VEC{S}_1\times\VEC{S}_2|$ of the opening angle, and the direction of the cross product is indicated by $\VEC{u}_s(\theta,\phi) = \VEC{S}_1\times\VEC{S}_2/|\VEC{S}_1\times\VEC{S}_2|$.
Considering only isotropic and chiral interactions, for these structures the total magnetic energy per trimer atom is (c.f.\ Eq.~\eqref{eq:energy3trimer})
\begin{align}\label{eq:energy_trimer_sym_opening}
    \MC{E}_s(\theta,\phi) &= \alpha(\theta) \big(J_{12} + \alpha(\theta) (B_{12} + B_{123})\big) \nonumber\\
    &+ \beta(\theta) \VEC{u}_s(\theta,\phi) \cdot \big(\VEC{D}_{12} + {\alpha(\theta) (\VEC{C}_{12} + \VEC{C}_{123}')}\big) \nonumber\\
    &= \alpha(\theta) \widetilde{J}_{12}(\theta) + \beta(\theta) \VEC{u}_s(\theta,\phi) \cdot \widetilde{\VEC{D}}_{12}(\theta) \;.
\end{align}

The isotropic interactions combine into an effective interaction $\widetilde{J}_{12}(\theta)$, while the chiral interactions form the combined chiral interaction vector $\widetilde{\VEC{D}}_{12}(\theta)$.
Here $\VEC{C}_{123}' = (0,C_{123}^y,C_{123}^z)$, as for these magnetic structures $C_{123}^x$ does not contribute.
The chiral part of the energy can be split into an out-of-plane contribution,
\begin{equation}
    \MC{E}^z_s(\theta,\phi) = s\,\frac{\sqrt{3}}{2}\,\widetilde{D}_{12}^z(\theta)\sin^2\theta \;,
\end{equation}
and an in-plane contribution
\begin{align}
    \MC{E}^y_+(\theta,\phi) &= -\frac{\sqrt{3}}{2}\,\widetilde{D}_{12}^y(\theta) \cos\phi \sin(2\theta) \nonumber\\
    \MC{E}^y_-(\theta,\phi) &= 0\;.
\end{align}
This shows that the $z$-component distinguishes between the two planar N\'eel states shown in Fig.~\ref{fig:trimer_states}, favoring one or the other depending on its sign.
The $y$-component only results in an energy gain for the structures with an anticlockwise rotation of the spins, which is maximized for $\phi = \SI{0}{\degree}$ if $\widetilde{D}_{12}^y(\theta) \sin(2\theta) > 0$ and for $\phi = \SI{180}{\degree}$ otherwise.
A structure that would favor a clockwise rotation of the spins around the triangle might still be able to gain energy from the in-plane component of the effective chiral vector, if the opening angles are not fixed to be the same for all pairs of spins.

\begin{figure}[tb]
\begin{center}
	\includegraphics[width=\columnwidth]{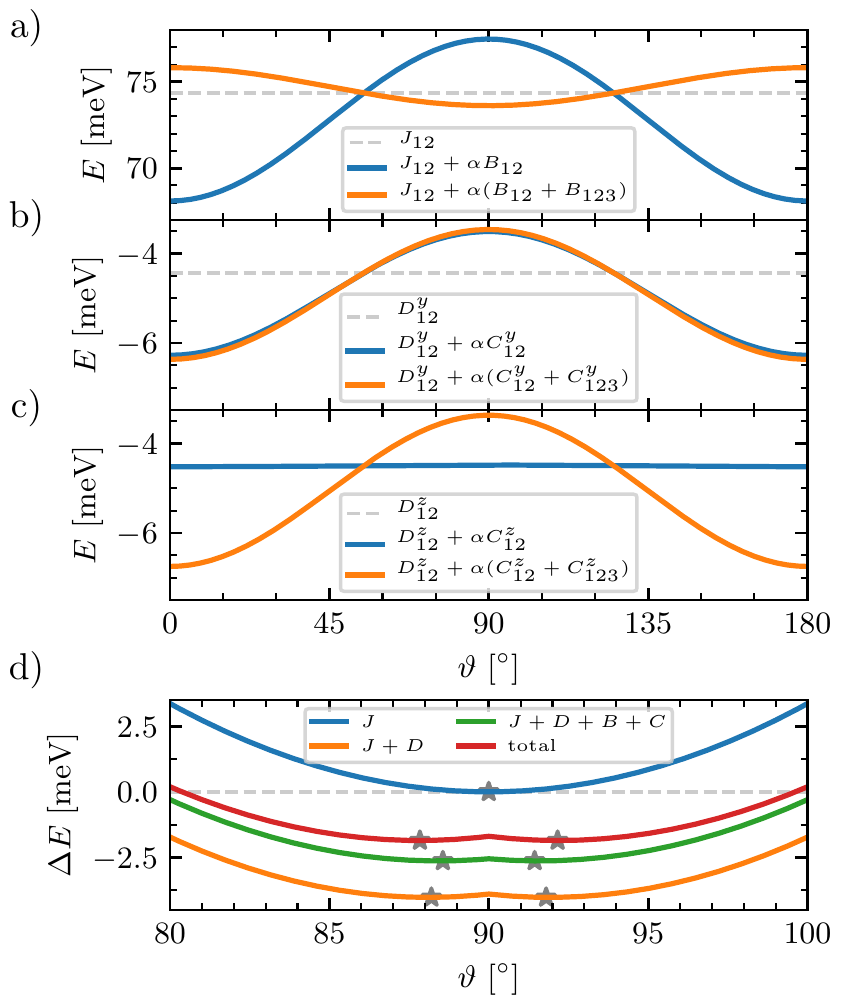}
	\caption{\label{fig:trimer_angular_dependence}
	Contributions to the energy of $C_\MR{3v}$-symmetric magnetic structures of Cr$_3$ on Pt(111).
	(a) Effective isotropic interaction $\widetilde{J}_{12}(\theta) = J_{12} + \alpha(\theta) (B_{12} + B_{123})$.
	(b) $y$-component and (c) $z$-component of the effective chiral interaction vector $\widetilde{\VEC{D}}_{12}(\theta) = \VEC{D}_{12} + \alpha(\theta) (\VEC{C}_{12} + \VEC{C}_{123}')$.
	(d) Total magnetic energy per trimer atom relative to the energy of the planar N\'eel state ($\theta = \SI{90}{\degree}$) given by $J_{12}$ alone.
	The analytic form of the energy due to the isotropic and chiral interactions is given in Eq.~\eqref{eq:energy_trimer_sym_opening}.
	For each $\theta$ we consider the value of $\phi$ that minimizes the energy.
	The last curve includes the contribution from the symmetric anisotropic interactions.
	The energy minima are indicated by the grey stars.
	}
\end{center}
\end{figure}

For these types of magnetic structures, the isotropic biquadratic and 3-site interactions contribute to the energy as $B_{12} + B_{123}$.
As seen from Table~\ref{tab:trimer_exchange_parameters}, these interactions have opposite signs for most trimers, cancelling out almost completely for Cr$_3$ on Pt(111) --- see Fig.~\ref{fig:trimer_angular_dependence}(a), Mn$_3$ on Re(0001) and Fe$_3$ on Au(111).
When $J_{12}$ is the dominant interaction, we can approximate $\alpha(\theta) \approx 1 - \frac{3}{2}\theta^2$ for $J_{12} < 0$ (canted ferromagnetic), from which we obtain that the isotropic energy increases with the tilt angle with the coefficient $J_{12} + 2 (B_{12} + B_{123})$, while for $J_{12} > 0$ we have instead $\alpha(\SI{90}{\degree}+\theta) \approx -\frac{1}{2} + \frac{3}{2}\theta^2$ (canted planar N\'eel), with the corresponding coefficient being $J_{12} - (B_{12} + B_{123})$.
Thus, the change in the energy from tilting the spins is influenced by the isotropic 4-spin interactions in different ways depending on the chosen reference configuration, and can be an indirect way of establishing their relative importance.
Overall, we see that it is important to include the different types of isotropic interactions, and not just the 2-site biquadratic interactions, when extending the spin model.

The chiral interactions lead to a canting of the magnetic structure with a well-defined chirality.
This is controlled by the effective chiral vector, which has contributions from the DMI, the CBI and the chiral 3-site interactions.
As with the isotropic interactions, the latter two interactions can either cooperate and compete, and as they are vector interactions this can happen in different ways for the different vector components.
The CBI and the chiral 3-site interaction for Cr$_3$ on Pt(111) have comparable magnitudes but are almost perpendicular to each other, combining into a vector $\VEC{C}_{12} + \VEC{C}_{123}'$ which is almost half of the DMI in magnitude.
However, the additional dependence on $\alpha(\theta)$ of the CBI and the chiral 3-site interaction conspires to cancel the enhancement of the DMI near $\theta = \SI{90}{\degree}$, instead leading to its weakening, as shown in Fig.~\ref{fig:trimer_angular_dependence}(b,c).
The different contributions to the energy are shown in Fig.~\ref{fig:trimer_angular_dependence}(d).
The strong antiferromagnetic $J_{12}$ favors the planar N\'eel state, which becomes canted due to the usual DMI --- the two minima near $\theta = \SI{90}{\degree}$ correspond to two canted N$_+$ states, with the spins tilting either towards the center of the trimer or away from it.
The chiral 4-spin interactions weaken the canting of the N\'eel state.
Finally, the contribution to the energy from the symmetric anisotropic interactions (not included in Eq.~\eqref{eq:energy_trimer_sym_opening}) modifies the canting angle, due to the tilted easy-axis of the single-site magnetic anisotropy.

While the previously discussed rotational symmetric magnetic structures showed the importance of the Moriya-like components of the chiral 3-site interaction, it did not give access to the states favoured by $C_{123}^x$.
Following Eq.~\eqref{eq:3site} and Fig.~\ref{fig:trianglechiral}, a typical term of the non-Moriya component is given by
\begin{equation}
    \MC{E}_{123}^x + \MC{E}_{213}^x = C_{123}^x \left(\VEC{S}_1 \times \VEC{S}_2\right)_x \left(\VEC{S}_2 - \VEC{S}_1\right) \cdot \VEC{S}_3 \;,
\end{equation}
while the same combination gives for the Moriya components
\begin{equation}
    \MC{E}_{123}^y + \MC{E}_{213}^y = C_{123}^y \left(\VEC{S}_1 \times \VEC{S}_2\right)_y \left(\VEC{S}_2 + \VEC{S}_1\right) \cdot \VEC{S}_3 \;,
\end{equation}
and similarly for $C_{123}^z$.
This exemplifies why the non-Moriya interaction is inoperative for magnetic structures where all the mutual angles are the same, which implies $\left(\VEC{S}_2 - \VEC{S}_1\right) \cdot \VEC{S}_3 = 0$, as we discussed in relation to Eq.~\eqref{eq:energy_trimer_sym_opening}.
However, if one started from an up-up-down state (see Fig.~\ref{fig:trimer_states}) then $\left(\VEC{S}_2 - \VEC{S}_1\right) \cdot \VEC{S}_3 = -2$, the non-Moriya interaction becomes active and would lead to a canting of this magnetic structure.

\begin{figure}[tb]
    \includegraphics[width=1.\columnwidth]{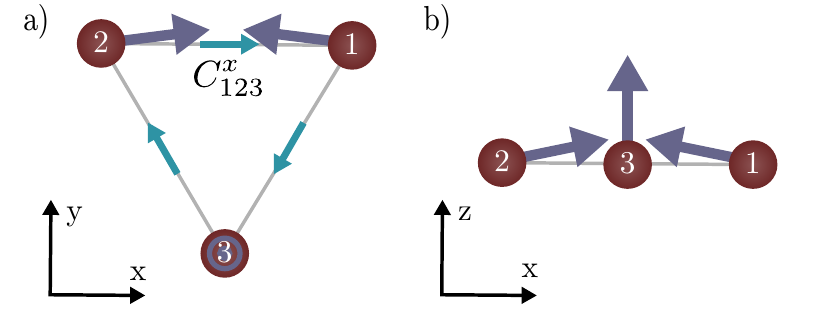}
    \caption{\label{fig:chiral_3site_groundstate}
    Magnetic structure favored by the non-Moriya component of the chiral 3-site interaction, $C_{123}^x$.
    a) Top view. b) Side view. Spherical coordinates: $\theta_1 = \theta_2 = \SI{78.6}{\degree}$, $\theta_3 = \SI{0}{\degree}$, $\phi_1 = \SI{173.0}{\degree}$ and $\phi_2 = \SI{7.0}{\degree}$.}
\end{figure}

One way of comparing the different chiral interactions acting on the trimer is to find what is the magnetic structure that is favored by each type of interaction on its own.
The out-of-plane DMI ($D_{12}^z$) favors the planar N\'eel state, selecting either N$_+$ or N$_-$ depending on its sign (see Fig.~\ref{fig:trimer_states}).
The in-plane DMI ($D_{12}^y$) favors a canted form of the N$_+$ state, with all spins having the same polar angle of $\theta = \SI{45}{\degree}$ and with the spins tilting either towards the center of the trimer or away from it.
The chiral biquadratic and the Moriya components of the chiral 3-site interactions favor similar states as the usual DMI.
The out-of-plane components ($C_{12}^z$ and $C_{123}^z$) also favor either N$_+$ or N$_-$, while the in-plane components ($C_{12}^y$ and $C_{123}^y$) favors a canted form of the N$_+$ state, but with a different value of the polar angle, $\theta = \SI{25.5}{\degree}$.
In contrast, the non-Moriya contribution favors a completely different state, shown in Fig.~\ref{fig:chiral_3site_groundstate}.
One of the spins stays normal to the plane of the trimer, while the other two cant symmetrically towards each other ($\phi_1 = \SI{173.0}{\degree}$ and $\phi_2 = \SI{7.0}{\degree}$) with a large opening angle with respect to the out-of-plane spin ($\theta_1 = \theta_2 = \SI{78.6}{\degree}$).

\subsection{Tetramers with $C_\MR{4v}$ symmetry}\label{sec:results_tetramers}
A tetramer of identical magnetic atoms arranged as a compact square on an fcc(001) surface has $C_\MR{4v}$ symmetry.
The chosen coordinate system, labelling of the magnetic atoms and mirror planes are shown in Fig.~\ref{fig:symmetries_illustration}b, and the symmetry matrices are described in Appendix~\ref{app:symm}.
The isotropic and chiral interactions for the tetramer are specified by Eqs.~\eqref{eq:2site}, \eqref{eq:3site} and \eqref{eq:4site}, with the detailed form for 4-site interactions given in Eq.~\eqref{eq:energy4tetramer}.
The reference interaction parameters are given in Table~\ref{tab:tetramer_exchange_parameters}, from which the full spin model can be parametrized by applying the $C_\MR{4v}$-symmetry operations.
The symmetric anisotropy parameters (see Eqs.~\eqref{eq:1site} and Eq.~\eqref{eq:2site} and Appendix~\ref{app:wizardry}) are not of our primary interest, but can be found in Table~\ref{tab:tetramer_symmetric_aniso_parameters}.

    \begin{table*}[htb]
        \centering
        \begin{ruledtabular}
            \renewcommand{\arraystretch}{1.1}
            \begin{tabular}{lrrrrrrrrrr}
            & $J_{12}$ & $J_{24}$ & $D_{12}^y$ & $D_{12}^z$ & $D_{24}$ & $B_{12}$ & $B_{24}$ & $C_{12}^y$ & $C_{12}^z$ & $C_{24}$ \\ \cline{1-11}
            Cr$_4$ &  29.09 &  11.69 &   0.42 &   1.22 &  -0.70 & -11.54 &  -0.58 &   2.34 &   0.66 &   0.11   \\
            Fe$_4$ &  -25.07 &  -2.44 &  -1.28 &  -0.40 &   4.35 &  -4.34 &  -0.20 &   0.98 &   0.52 &  -0.47
            \end{tabular}
            \vspace{-3.8pt}
            \begin{tabular}{lrrrrrrrrrrrrrrrr}
            & $B_{123}$ & $B_{124}$ & $C_{123}^x$ & $C_{123}^y$ & $ C_{123}^z$ & $C_{124}^x$ &  $C_{124}^y$ & $C_{124}^z$ & $C_{421}^x$ & $C_{421}^y$ & $C_{421}^z$ & $B_{1234}$ & $B_{1324}$ & $C_{1234}^y$ & $C_{1234}^z$ & $C_{2413}$ \\ \hline
            Cr$_4$ &   -2.69 &  -0.82 &  -0.89 &   0.81 &  -1.91 &   0.40 &   0.61 &   0.09 &  -0.31 &  -0.26 &   0.32 &  -1.15 &   1.26 &   0.87 &   0.95 &   1.19  \\ %\cline{2-13}
            Fe$_4$ &  0.20 &   1.09 &  -0.26 &  -0.37 &  -1.57 &   0.04 &  -0.72 &  -0.52 &  -0.16 &  -0.26 &   0.22 &   0.15 &   1.10 &  -0.86 &  -0.28 &  -0.60 
            \end{tabular}
        \end{ruledtabular}
    	\caption{
    	Magnetic interactions in different compact tetramers deposited on the Pt(100) surface in units of [meV].
    	We give the reference parameters for all isotropic and chiral 2-site, 3-site and 4-site interactions.
    	The full set of interactions can be obtained from the shown ones by applying the $C_\MR{4v}$-symmetry operations to the tetramer, see Section~\ref{sec:c4v}.
    	}
    	\label{tab:tetramer_exchange_parameters}
    \end{table*}

We first discuss possible scenarios considering the isotropic interactions alone.
The dominant interaction is the nearest-neighbor interaction $J_{12}$.
$J_{12} < 0$ favors a ferromagnetic alignment, while $J_{12} > 0$ favors the collinear up-down-up-down antiferromagnetic state (going around the tetramer).
As the tetramer allows for next-nearest-neighbor interactions, $J_{13}$ could in principle compete with $J_{12}$ and stabilize an up-up-down-down state, but this requires $J_{13} \geq |J_{12}|$, a condition far from being satisfied in our tetramers.
The isotropic biquadratic and 4-site interactions contribute to the energy in exactly the same way for these three collinear states, $\MC{E} = 4 \left(B_{12} + B_{1234}\right) + 2 \left(B_{13} + B_{1324}\right)$, and so do not favor any of them.
The isotropic 3-site interactions do distinguish between the three collinear states: $B_{123}$ distinguishes the up-up-down-down state from the other two, while $B_{124}$ distinguishes all three of them.
However, the magnitude of these isotropic 4-spin interactions for our tetramers is not strong enough to modify the collinear state favored by the nearest-neighbor interaction $J_{12}$.

In order to easily compare the isotropic and chiral contributions to the magnetic energy of the tetramer, we now consider a family of magnetic structures with fourfold rotational symmetry.
We parametrize the spins as in Eq.~\eqref{eq:magstruct}, but now setting $\phi_1 = \phi + \SI{45}{\degree}$, $\phi_2 = \phi_1 + s\,\SI{90}{\degree}$, $\phi_3 = \phi_1 + s\,\SI{180}{\degree}$ and $\phi_4 = \phi_1 + s\,\SI{270}{\degree}$ with $s=\{+,-\}$.
As before, $\theta = \SI{0}{\degree}$ or \SI{180}{\degree} corresponds to the ferromagnetic state.
However, $\theta = \SI{90}{\degree}$ is not the up-down-up-down state, but a planar noncollinear state with all spins perpendicular to their nearest-neighbors.
We define $\alpha(\theta) = \VEC{S}_1\cdot\VEC{S}_2$ and $\beta(\theta) = |\VEC{S}_1\times\VEC{S}_2|$, with the direction of the cross product indicated by $\VEC{u}_s(\theta,\phi) = \VEC{S}_1\times\VEC{S}_2/|\VEC{S}_1\times\VEC{S}_2|$, which together characterize the openings of nearest-neighbor spins.
We also define $\alpha'(\theta) = \VEC{S}_2\cdot\VEC{S}_4$, $\beta'(\theta) = |\VEC{S}_2\times\VEC{S}_4|$ and $\VEC{u}_s'(\theta,\phi) = \VEC{S}_2\times\VEC{S}_4/|\VEC{S}_2\times\VEC{S}_4|$, characterizing the openings of next-nearest-neighbor spins.
Considering only isotropic and chiral interactions, for these structures the total magnetic energy per tetramer atom is (c.f.\ Eq.~\eqref{eq:energy4tetramer})
\begin{align}\label{eq:energy_tetramer_sym_opening}
    \MC{E}_s(\theta,\phi) &= \alpha \big(J_{12} + \alpha (B_{12} + B_{123} + B_{1234}) + \alpha' B_{124}\big) \nonumber\\
    &+ \frac{\alpha'}{2} \big(J_{24} + \alpha' (B_{24} + B_{1324}) + 2 \alpha B_{124}\big) \nonumber\\
    &+ \beta \VEC{u}_s \cdot \big(\VEC{D}_{12} + \alpha (\VEC{C}_{12} + \VEC{C}_{123}' + \VEC{C}_{1234}) + \alpha'\VEC{C}_{124}'\big) \nonumber\\
    &+ \frac{\beta'}{2} \VEC{u}_s' \cdot \big(\VEC{D}_{24} + \alpha' (\VEC{C}_{24} + \VEC{C}_{2413}) - \alpha\VEC{C}_{421}'\big) \nonumber\\
    &= \alpha \widetilde{J}_{12} + \frac{\alpha'}{2} \widetilde{J}_{24}
    + \beta \VEC{u}_s \cdot \widetilde{\VEC{D}}_{12} + \frac{\beta'}{2} \VEC{u}_s' \cdot \widetilde{\VEC{D}}_{24} \;.
\end{align}
Here $\VEC{C}_{123}' = (0,C_{123}^y,C_{123}^z)$ and likewise for $\VEC{C}_{124}'$.
Lastly, $\VEC{C}_{421}' = C_{421}' \left(\frac{1}{\sqrt{2}},\frac{1}{\sqrt{2}},0\right)$ with $C_{421}' = \sqrt{2}\left(C_{421}^x + C_{421}^y\right)$, so only the net component which is collinear with $\VEC{D}_{24}$ contributes.
Hence the components $C_{123}^x$, $C_{124}^x$ and $C_{421}^z$ do not contribute to the energy of these magnetic structures.
As in the case of the trimer, we can interpret the different contributions to the energy as if arising from effective interactions $\widetilde{J}_{ij}$ and $\widetilde{\VEC{D}}_{ij}$.

The chiral part of the energy can be split into an out-of-plane contribution,
\begin{equation}
    \MC{E}^z_s(\theta,\phi) = s\,\widetilde{D}_{12}^z(\theta)\sin^2\theta \;,
\end{equation}
and an in-plane contribution
\begin{align}
    \MC{E}^y_+(\theta,\phi) &= -\left(\frac{1}{\sqrt{2}}\widetilde{D}_{12}^y(\theta)
    + \frac{1}{2} \widetilde{D}_{24}(\theta)\right) \cos\phi \sin(2\theta) \;, \nonumber\\
    \MC{E}^y_-(\theta,\phi) &= 0\;.
\end{align}
As for the trimer, only one chirality can gain energy from the in-plane components of the effective DMI.

We now focus on the Fe tetramer, for which $J_{12} = \SI{-25}{\milli\electronvolt}$ and $J_{24} = \SI{-2.4}{\milli\electronvolt}$ favor a ferromagnetic structure, and so an opening due to chiral interactions is expected.
For a small opening, we can set $\alpha(\theta) \approx 1 - \theta^2$ and $\alpha'(\theta) \approx  1 - 2 \theta^2$, so that the second derivative of the isotropic energy has the coefficients $J_{12} + 2\left(B_{12} + B_{123} + B_{1234}\right) + 3 B_{124} = \SI{-30}{\milli\electronvolt}$ and $J_{24} + 2\left(B_{24} + B_{1324}\right) + 3 B_{124} = \SI{2.6}{\milli\electronvolt}$, showing that the isotropic 4-spin interactions make a very important contribution and can even reverse the effective sign of the interaction.
The nearest-neighbor DMI $|\VEC{D}_{12}| = \SI{1.3}{\milli\electronvolt}$ is strongly enhanced close to the ferromagnetic state to $|\widetilde{\VEC{D}}_{12}| = \SI{3.2}{\milli\electronvolt}$, while the next-nearest-neighbor DMI $D_{24} = \SI{4.4}{\milli\electronvolt}$ is slightly weakened to $\widetilde{D}_{24} = \SI{3.7}{\milli\electronvolt}$.
When all interactions are taken together, the ground state of the Fe tetramer on Pt(001) is found to be a ferromagnetic structure almost perpendicular to the plane of the surface, with a symmetric canting of all spins ($s = +$) away from the center of the tetramer, with $\theta = \SI{8}{\degree}$.
The main origin of this canting are the chiral interactions across the diagonals of the tetramer.

\begin{figure}[tb]
    \includegraphics[width=1.\columnwidth]{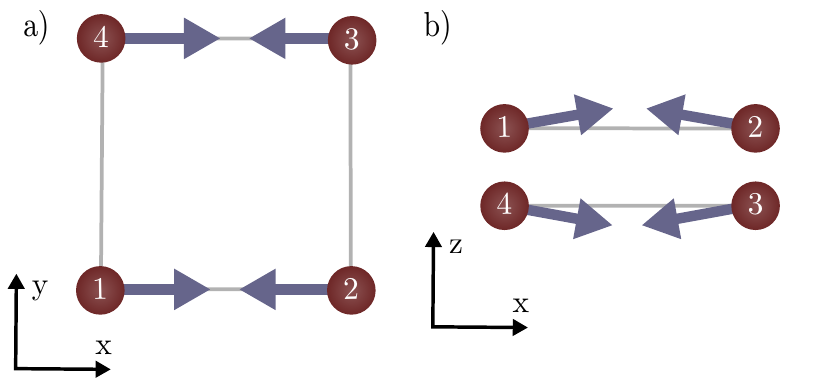}
    \caption{\label{fig:Cr_tetramer_groundstate}
    Magnetic structure of the Cr tetramer obtained considering the full set of magnetic interactions.
    a) Top view. b) Side view. Spherical coordinates: $\theta_1 = \theta_2 = \SI{92.3}{\degree}$, $\theta_3 = \theta_4 = \SI{87.7}{\degree}$, $\phi_1 = \phi_4 = \SI{0.1}{\degree}$, and $\phi_2 = \phi_3 = \SI{179.9}{\degree}$.}
\end{figure}

The Cr tetramer has strong antiferromagnetic interactions favoring the collinear antiferromagnetic up-down-up-down state.
By setting $\theta_1 = \theta_3 = \theta$ and $\theta_2 = \theta_4 = \SI{180}{\degree} - \theta$, one can still use Eq.~\eqref{eq:energy_tetramer_sym_opening} but with $\alpha \rightarrow -\alpha$ and taking care of the change in handedness of some cross products, which leads to
\begin{align}
    \MC{E}^y_+(\theta,\phi) &= 0 \;, \nonumber\\
    \MC{E}^y_-(\theta,\phi) &= \left(\frac{1}{\sqrt{2}}\widetilde{D}_{12}^y(\theta)
    - \frac{1}{2} \widetilde{D}_{24}(\theta)\right) \cos\phi \sin(2\theta) \;.
\end{align}
The out-of-plane contribution from the chiral interactions is unchanged.
The strongest interactions are $J_{12} = \SI{29}{\milli\electronvolt}$ and $J_{24} = \SI{12}{\milli\electronvolt}$.
For a small opening, we can set $\alpha(\theta) \approx -1 + \theta^2$ and $\alpha'(\theta) \approx 1 - 2\theta^2$, so that the second derivative of the isotropic energy has the coefficients $J_{12} - 2\left(B_{12} + B_{123} + B_{1234}\right) + 3 B_{124} = \SI{57}{\milli\electronvolt}$ and $J_{24} + 2\left(B_{24} + B_{1324}\right) - 3 B_{124} = \SI{16}{\milli\electronvolt}$, showing a very strong contribution from the isotropic 4-spin interactions to the nearest-neighbor part.
The nearest-neighbor DMI $|\VEC{D}_{12}| = \SI{1.3}{\milli\electronvolt}$ is strongly enhanced close to the ferromagnetic state to $|\widetilde{\VEC{D}}_{12}| = \SI{3.4}{\milli\electronvolt}$, while the next-nearest-neighbor DMI $D_{24} = \SI{-0.7}{\milli\electronvolt}$ is reduced to almost zero.
When all interactions are taken together, the ground state of the Cr tetramer on Pt(001) is found to be a slightly canted up-down-up-down state almost collinear with either the $x$- or $y$-directions, due to the symmetric anisotropic interactions, and the spins tilt away from the $xy$-plane by $\Delta\theta = \pm\SI{2.3}{\degree}$.
This magnetic structure is illustrated in Fig.~\ref{fig:Cr_tetramer_groundstate}.

\section{Relation to other works}
We now relate our findings to other works addressing magnetic interactions in very disparate systems, highlighting common ground and clarifying several aspects concerning the multi-site interactions.

First we would like to mention that Bornemann et al.\cite{Bornemann2012} presented an extensive survey of the magnetic properties of diverse clusters of Fe, Co and Ni on Ir(111), Pt(111) and Au(111), using the infinitesimal rotation method based on the ferromagnetic state as reference.
However, direct comparison of our data in Table~\ref{tab:trimer_exchange_parameters} with their reported values is difficult due to their neglect of structural relaxations and various other computational differences, so we will not attempt this here.
We have previously calculated the magnetic exchange interactions for Fe trimers on Pt(111), in connection to scanning tunneling microscopy experiments\cite{Hermenau2017,Hermenau2019}.
In Ref.~\onlinecite{Hermenau2017} we employed the infinitesimal rotation method for the 2-spin interactions with a correction due to the spin polarizability of Pt, obtaining $J_{12} = \SI{-54}{\milli\electronvolt}$, $D_{12}^y = \SI{1.3}{\milli\electronvolt}$, and $D_{12}^z = \SI{-0.7}{\milli\electronvolt}$.
As discussed in Sec.~\ref{sec:results_trimers}, we should compare $J_{12}$ from the infinitesimal rotation method with $J_{12} + 2 (B_{12} + B_{123}) =  \SI{-44}{\milli\electronvolt}$, and the DMI to $\widetilde{D}_{12}^y = \SI{4.9}{\milli\electronvolt}$ and $\widetilde{D}_{12}^z = \SI{-0.4}{\milli\electronvolt}$, as defined in Eq.~\eqref{eq:energy_trimer_sym_opening}.
We believe that the discrepancies can be explained by the different treatment of the potential: Ref.~\onlinecite{Hermenau2017} employed the atomic sphere approximation while our present work makes no shape approximation to the potential.

To the best of our knowledge, there is only one previous work addressing isotropic 4-spin interactions in magnetic clusters, Ref.~\onlinecite{Antal2008}, using a different computational geometry and approximations, which once again cautions against quantitative comparisons.
Ref.~\onlinecite{Antal2008} reports for the isotropic 4-spin interactions of Cr$_3$ on Au(111) $B_{12} = \SI{-4.4}{\milli\electronvolt}$ vs.\ our $B_{12} = \SI{-5.1}{\milli\electronvolt}$, and $B_{123} = \SI{7.1}{\milli\electronvolt}$ vs.\ our $B_{123} = \SI{8.1}{\milli\electronvolt}$.
Both values are in fair agreement with ours, also concerning the opposite sign of these interactions.
They also report $J_{12} = \SI{145}{\milli\electronvolt}$ vs.\ our $J_{12} = \SI{88}{\milli\electronvolt}$, and $|\VEC{D}_{12}| = \SI{1.8}{\milli\electronvolt}$ vs.\ our $|\VEC{D}_{12}| = \SI{4.3}{\milli\electronvolt}$, both quite different from ours, so the numerical agreement at the level of the 4-spin interactions might be fortuitous.

Fe chains on Re(0001) were experimentally found to have a short-period spin-spiral ground state\cite{Kim2018b,schneider2020controlling}.
While exploring the magnetic properties of this system with DFT calculations, L\'aszl\'offy et al.\cite{Laszloffy2019} found an inconsistency between the noncollinear spin structure obtained via magnetic force theorem calculations and the one obtained with an atomistic spin model containing only 2-spin interactions.
While trying to understand the origin of this inconsistency, they introduced on phenomenological grounds chiral multi-site interactions.
Our systematic procedure for constructing chiral multi-site interactions fully justifies the phenomenological forms proposed by these authors.
We note, however, that we did not recover the aforementioned inconsistency when revisiting the same system without any shape approximation to the potential, instead finding a spin structure which is consistent with the experimental one\cite{schneider2020controlling}.

Grytsiuk et al.\cite{Grytsiuk2020} investigated theoretically the complex magnetism of MnGe\cite{Tanigaki2015,Fujishiro2019}.
Given that the Mn atoms in MnGe are coordinated in triangular plaquettes, the authors proposed so-called topological-chiral interactions which are built upon the scalar spin chirality of the three spins forming one such triangle, $\chi_{123} = \VEC{S}_1 \cdot \left(\VEC{S}_2 \times \VEC{S}_3\right)$.
The chiral-chiral interaction is a 6-spin 3-site interaction $\kappa_{123}^\MR{CCI} \left(\chi_{123}\right)^2$ that does not require SOC, while the spin chiral interaction is a 4-spin 3-site interaction built of terms such as $\left(\VEC{C}_{123}^\MR{SCI}\cdot\VEC{S}_1\right)\chi_{123}$ and is driven by SOC.
DFT calculations showed that both types of interactions are quite strong in MnGe.
Given our proposed classification of multi-site interactions into isotropic (no SOC required) and chiral (SOC required) interactions, the question arises as to how these topological-chiral interactions fit this classification.
In fact, we show in Appendix~\ref{app:sci} that the chiral-chiral interaction can be expressed solely using dot products of the involved spins, so it is of the generic form of the isotropic interactions, and that the spin-chiral interaction can be rewritten using combinations of dot products and cross products, thus being covered by the chiral 4-spin interactions that we discuss in the present work.
However, interactions built solely out of the scalar spin chirality cannot reproduce all the different types of interactions that we uncovered in the present work (see Appendix ~\ref{app:sci}), and for a complete spin model our systematic forms for the interactions should be used.

Cardias et al.\cite{Cardias2020} recently submitted a preprint titled `Dzyaloshinskii-Moriya interaction in absence of spin-orbit coupling'.
Our microscopic model and heuristic arguments concerning the forms of the magnetic interactions unequivocally show that without SOC the interactions are isotropic, no cross products of spins appear, and without cross products there are no DMI-like chiral interactions.
We are convinced that these puzzling findings can be explained using the generalized atomistic spin model that we discussed in this work, by identifying the type of isotropic multi-spin and/or multi-site interactions that lead to the obtained angular dependence of the energy or its first derivative.

As we mentioned in the Introduction, several works have proposed ways of extending the infinitesimal rotation method to 4-spin interactions\cite{Lounis2010,Szilva2013,Grytsiuk2020,Mankovsky2019}.
In particular, Ref.~\onlinecite{Mankovsky2019} discusses both isotropic and chiral multi-site interactions, and derives the corresponding expressions for their calculation in a DFT context.
They showed that the chiral 4-spin 3-site interactions do not vanish for centrosymmetric systems with the example of bcc Fe, in full agreement with our symmetry analysis of these interactions.
The authors also introduce a chiral 3-spin 3-site interaction which is defined through the scalar spin chirality $\chi_{123}$.
This kind of interaction is not time-reversal-invariant unless the interaction coefficient also changes sign, and so we believe that the corresponding calculations need to be reinterpreted in a way that complies with time-reversal symmetry.

\section{Conclusions}
In this work, we presented a comprehensive framework for isotropic and chiral multi-site interactions, along with systematic calculations of these interactions for several magnetic trimers and tetramers.

First, we imposed the general requirement of time-reversal invariance of the magnetic energy on the possible form of the interactions, and gave simple heuristic arguments that can be used to obtain the form of the multi-site interactions.
We thus arrived at a generalized atomistic spin model containing 2-spin and 4-spin interactions that couple up to four distinct magnetic sites.
Next we demanded that our interactions comply with the crystallographic point group symmetry, with the concrete examples of $C_\MR{3v}$ (trimers) and $C_\MR{4v}$ (tetramers).
Contrary to the 1-site and 2-site interactions, those based on three or four sites are much less constrained by the point group symmetry operations.
For instance, while the 2-site Dzyaloshinskii-Moriya interaction vanishes if there is an inversion center in the middle of those two sites, this is not the case for the chiral 3- and 4-site interactions.
We also found that the respective chiral interaction vectors can have components which are forbidden for the DMI due to Moriya’s rules.
The chiral multi-site interactions do comply with a generalized Moriya rule: If all sites connected by the interaction lie in the same mirror plane, the chiral interaction vector must be perpendicular to this plane.

After outlining our global mapping scheme from DFT calculations to a target spin model, we presented our results on a series of homoatomic trimers and tetramers on several surfaces with strong spin-orbit coupling.
While in most-cases the dominant interaction is the familiar isotropic Heisenberg exchange, this is not so for the trimers on the Re(0001) surface.
For the trimers, we found that the isotropic biquadratic and 3-site interactions tend to counteract each other, while the chiral biquadratic and 3-site interactions more easily combine due to their vector nature, supporting or hindering the DMI depending on the magnetic structure and on the type of atoms forming the trimer.
We also discussed the magnetic structure favored by the non-Moriya component of the chiral 3-site interactions.
For the tetramers on Pt(001), the isotropic 4-spin interactions were found to cooperate and have a strong contribution for the Cr case, while the chiral 4-spin interactions dominate over the DMI, playing a leading role in the canting of the ground state magnetic structures of the tetramers.
Lastly, we briefly addressed recent proposals for multi-site interactions and placed them into the context of our work.

We believe that our work is a timely contribution to the growing research activity on materials with complex magnetic structures, such as multiple-$\VEC{Q}$ states\cite{Kurz2001,Heinze2011,Al-Zubi2011,Kroenlein2018,Romming2018,Spethmann2020}, for which the role of the chiral multi-site interactions is yet to be explored.
Our detailed exposition of the symmetry properties of the multi-site interactions, as well as concrete examples for their enumeration, should clarify the bookkeeping which is essential to properly account for all possible ways of combining a set of sites with a given type of magnetic interaction.
Our example systems, trimers and tetramers, are ubiquitous building blocks (triangles and squares, respectively) of many lattices, which will help transfer our findings to extended systems, from a single layer to bulk magnets.

\begin{acknowledgments}
 This work was supported by the European Research Council (ERC) under the European Union's Horizon 2020 research and innovation program (ERC-consolidator grant 681405 -- DYNASORE). 
 The authors gratefully acknowledge the computing time granted through JARA-HPC on the supercomputer JURECA at the Forschungszentrum J\"ulich \cite{jureca}.
\end{acknowledgments}

\begin{appendix}

\section{Symmetric anisotropy matrices}\label{app:wizardry}
Here we briefly explain the rationale being the form of the 1-site and 2-site anisotropy matrices used in this work.
Consider the following real symmetric matrix:
\begin{equation}
\renewcommand{\arraystretch}{1.2}
    A = \begin{pmatrix} A^{xx} & A^{xy} & A^{xz} \\ A^{xy} & A^{yy} & A^{yz} \\  A^{xz} & A^{yz} & A^{zz} \end{pmatrix} \;.
\end{equation}
By solving the eigensystem $A\,\VEC{u}_n = \lambda_n \VEC{u}_n$ where $\lambda_n$ are the eigenvalues and $\VEC{u}_n$ are the normalized eigenvectors ($\VEC{u}_n\cdot\VEC{u}_n = 1$) of the matrix $A$, one can write
\begin{equation}
    A = \lambda_1 \VEC{u}_1 \otimes \VEC{u}_1 + \lambda_2 \VEC{u}_2 \otimes \VEC{u}_2 + \lambda_3 \VEC{u}_3 \otimes \VEC{u}_3 \;.
\end{equation}
This defines an ellipsoid with principal axes of length given by the eigenvalues and orientation given by the eigenvectors.
As an example, consider the mirror symmetry $\MC{M} = 1 - 2\,\VEC{n}\otimes\VEC{n}$, where $\VEC{n}$ is the unit normal to the mirror plane.
If the anisotropic interaction is invariant under this mirror symmetry, $\MC{M} A \MC{M} = A$, then $\VEC{n}$ must be one of the eigenvectors of $A$.
        
Consider now the symmetric anisotropic interactions.
For $\sum_{\alpha,\beta} K_i^{\alpha\beta} S_i^\alpha S_i^\beta$ we have an on-site anisotropy matrix, for which only the energy differences when the spin is aligned with each of the principal axes is meaningful.
We can thus set one of its eigenvalues to zero, and we do this with the one invariant under the mirror plane, say $\lambda_3 = 0$ if $\VEC{n} = \VEC{u}_3$, which leaves a finite two-dimensional subspace $K_i = \lambda_1 \VEC{u}_1 \otimes \VEC{u}_1 + \lambda_2 \VEC{u}_2 \otimes \VEC{u}_2$.
This corresponds to the matrix $K_i$ having three independent parameters, which only reduce to two if the coordinate axes are aligned with the eigenvectors of this matrix.
If $i \neq j$ we have a symmetric anisotropic exchange matrix, and we cannot a priori set any eigenvalue to zero.
We can however rewrite $\sum_{\alpha,\beta} J_{ij}^{\alpha\beta} S_i^\alpha S_j^\beta = J_{ij}\,\VEC{S}_i \cdot \VEC{S}_j + \sum_{\alpha,\beta} \Delta J_{ij}^{\alpha\beta} S_i^\alpha S_j^\beta$.
Now the role of the anisotropy is isolated in the matrix $\Delta J_{ij}$, and we can choose to set to zero the eigenvalue that is invariant under the mirror symmetry, if the symmetry applies.
        
First we discuss the trimer with $C_\MR{3v}$ symmetry, see Fig.~\ref{fig:symmetries_illustration}a.
We take atom 3 as reference, for which $\MC{M}_3 K_3 \MC{M}_3 = K_3$ can be used to impose the form
\begin{equation}
    \renewcommand{\arraystretch}{1.2}\label{eq:konc3v}
    K_3 = \begin{pmatrix} 0 & 0 & 0 \\ 0 & K^{yy}_3 & K^{yz}_3 \\ 0 & K^{yz}_3 & K^{zz}_3 \end{pmatrix} \;.
\end{equation}
From this matrix the remaining ones are generated by $K_1 = \MC{R}_+ K_3 \MC{R}_-$ and $K_2 = \MC{R}_- K_3 \MC{R}_+$.
The symmetric anisotropic exchange is simplest to specify for atoms 1 and 2.
The mirror symmetry replaces $\VEC{S}_1 \rightarrow \MC{M}_3 \VEC{S}_2$ and $\VEC{S}_2 \rightarrow \MC{M}_3 \VEC{S}_1$, which leads to $\Delta J_{12} = \MC{M}_3 \Delta J_{12} \MC{M}_3$ (see Eq.~\eqref{eq:2site}).
In perfect analogy with the on-site anisotropy, we can thus write
\begin{equation}\label{eq:jijc3v}
    \renewcommand{\arraystretch}{1.2}
    \Delta J_{12} = \begin{pmatrix} 0 & 0 & 0 \\ 0 & \Delta J^{yy}_{12} & \Delta J^{yz}_{12} \\ 0 & \Delta J^{yz}_{12} & \Delta J^{zz}_{12} \end{pmatrix} \;.
\end{equation}
The remaining matrices can be obtained by rotation, $\Delta J_{23} = \MC{R}_+ \Delta J_{12} \MC{R}_-$, $\Delta J_{31} = \MC{R}_- \Delta J_{12} \MC{R}_+$, etc.
The parameters of the symmetric anisotropic matrices for the trimers are shown in Table \ref{tab:trimer_symmetric_aniso_parameters}.

    \begin{table}[!t]
        \centering
        \begin{ruledtabular}
            \begin{tabular}{lrrrrrrr}
            Surface & System &  $K_3^{yy}$ & $K_3^{zz}$ & $K_3^{yz}$ & $\Delta J_{12}^{yy}$ & $\Delta J_{12}^{zz}$ & $\Delta J_{12}^{yz}$ \\ \hline
                \multirow{4}{*}{Pt(111)}   & Cr$_3$ &   1.02 &   2.82 &  -1.08 &  -0.04 &  -0.17 &   0.29    \\%\cline{2-8}
                                           & Mn$_3$ &  -2.68 &  -1.15 &   1.00 &   0.05 &  -0.72 &   0.29   \\%\cline{2-8}
                                           & Fe$_3$ &  -0.56 &  -0.31 &   2.47 &   0.68 &   0.10 &  -0.65   \\%\cline{2-8}
                                           & Co$_3$ &  -0.74 &  -4.33 &   0.40 &   0.14 &  -0.07 &  -0.47   \\\hline
                 \multirow{4}{*}{Re(0001)} & Cr$_3$ &   0.97 &  -3.96 &  -1.21 &  -0.23 &   1.43 &  -0.06   \\%\cline{2-8}
                                           & Mn$_3$ &   1.48 &  -5.71 &  -1.82 &  -1.16 &   1.04 &   0.18   \\%\cline{2-8}
                                           & Fe$_3$ &   2.18 &   2.59 &  -0.89 &  -0.33 &  -0.65 &   0.05   \\%\cline{2-8}
                                           & Co$_3$ &  -0.95 &   2.51 &   0.40 &   0.00 &   0.33 &  -0.27   \\\hline
                \multirow{4}{*}{Au(111)}   & Cr$_3$ &   0.13 &   0.38 &  -0.38 &   0.03 &  -0.08 &   0.09   \\%\cline{2-8}
                                           & Mn$_3$ &  -0.03 &   1.29 &   0.78 &   0.39 &  -0.00 &  -0.45   \\%\cline{2-8}
                                           & Fe$_3$ &  -0.27 &  -1.41 &   0.46 &  -0.20 &   0.26 &  -0.10   \\%\cline{2-8}
                                           & Co$_3$ &  -2.25 &  -4.72 &  -1.81 &  -0.26 &   0.49 &   0.31   \\
             \end{tabular}
        \end{ruledtabular}
    	\caption{
    	Parameters of the symmetric anisotropic matrices for the compact top-stacked trimers in units of [meV].
    	}
    	\label{tab:trimer_symmetric_aniso_parameters}
    \end{table}

    \begin{table}[!t]
        \centering
        \begin{ruledtabular}
            \begin{tabular}{crrrrrrrr}
            System & $K_1^{xx}$ & $K_1^{xz}$ & $K_1^{zz}$ & $\Delta J_{12}^{yy}$ & $\Delta J_{12}^{zz}$ & $\Delta J_{12}^{yz}$ & $\Delta J_{24}^{xx}$ & $\Delta J_{24}^{zz}$ \\
            Cr$_4$ & 0.04 &   0.10 &   2.44 &   -0.29 &  -1.41 &  -0.09 &   0.84 &   2.75 \\
            Fe$_4$ & 0.62 &   0.14 &  -0.66 &   -0.65 &   0.35 &   0.33 &   0.08 &  -2.49 \\
            \end{tabular}
        \end{ruledtabular}
    	\caption{
    	Parameters of the symmetric anisotropic matrices in different compact tetramers deposited on the Pt(100) surface in units of [meV].
    	}
    	\label{tab:tetramer_symmetric_aniso_parameters}
    \end{table}

Now we discuss the tetramer with $C_\MR{4v}$ symmetry, see Fig.~\ref{fig:symmetries_illustration}b.
The mirror planes $\MC{M}_\pm$ determine the form of the on-site anisotropy matrices.
These have three independent parameters as for the trimer, but due to the mirror planes not being aligned with the cartesian axes the matrices seem a bit more complicated.
We take atom 3 as reference, for which $\MC{M}_+ K_3 \MC{M}_+ = K_3$ can be used to impose the form
\begin{equation}\label{eq:konc4v}
    \renewcommand{\arraystretch}{1.2}
    K_3 = \begin{pmatrix} K^{xx}_3 & K^{xx}_3 & K^{xz}_3 \\ K^{xx}_3 & K^{xx}_3 & K^{xz}_3 \\ K^{xz}_3 & K^{xz}_3 & K^{zz}_3 \end{pmatrix} \;,
\end{equation}
with the other matrices being given by $K_1 = \MC{M}_- K_3 \MC{M}_-$, $K_2 = \MC{M}_y K_3 \MC{M}_y$ and $K_4 = \MC{M}_x K_3 \MC{M}_x$.
As $\Delta J_{12} = \MC{M}_x \Delta J_{12} \MC{M}_x$, the symmetric exchange matrix for nearest-neighbors $\Delta J_{12}$ has the same form as given for the trimer in Eq.~\eqref{eq:jijc3v}, and going around the edges of the tetramer we have $\Delta J_{14} = \MC{M}_+ \Delta J_{12} \MC{M}_+$,  $\Delta J_{43} = \MC{M}_y\Delta J_{12}\MC{M}_y$ and $\Delta J_{32} = \MC{M}_- \Delta J_{12} \MC{M}_-$.
For the next-nearest-neighbors we take atoms 1 and 3 as reference, for which $\Delta J_{13} = \MC{M}_+ \Delta J_{13} \MC{M}_+$ and $\Delta J_{13} = \MC{M}_- \Delta J_{13} \MC{M}_-$, leading to:
\begin{equation}\label{eq:jijc4v}
    \renewcommand{\arraystretch}{1.2}
    \Delta J_{13} = \begin{pmatrix} \Delta J^{xx}_{13} & \Delta J^{xx}_{13} & 0 \\
    \Delta J^{xx}_{13} & \Delta J^{xx}_{13} & 0 \\
    0 & 0 & \Delta J^{zz}_{13} \end{pmatrix} \;.
\end{equation}
Lastly, $\Delta J_{24} = \MC{M}_x \Delta J_{13} \MC{M}_x = \MC{M}_y \Delta J_{13} \MC{M}_y$.
The parameters of the symmetric anisotropic matrices for the tetramers are shown in Table \ref{tab:tetramer_symmetric_aniso_parameters}.

\section{Equivalent forms of the chiral 4-spin interactions and the isotropic 6-spin interaction}\label{app:sci}
First we show how to transform the so-called chiral-chiral interaction\cite{Grytsiuk2020}, $\kappa_{123}^\MR{CCI} \left(\chi_{123}\right)^2$, into combinations of dot products.
The scalar chirality can be expressed as the determinant of a matrix with the spins either as rows or as columns,
\begin{equation}
    \renewcommand{\arraystretch}{1.2}
    \chi_{123} = \det \begin{pmatrix} S_1^x & S_1^y & S_1^z \\ S_2^x & S_2^y & S_2^z \\ S_3^x & S_3^y & S_3^z \end{pmatrix}
     = \det \begin{pmatrix} S_1^x & S_2^x & S_3^x \\ S_1^y & S_2^y & S_3^y \\ S_1^z & S_2^z & S_3^z \end{pmatrix} \;.
\end{equation}
We can then write, using $(\det A)(\det B) = \det(AB)$,
\begin{align}
    \left(\chi_{123}\right)^2 &= \det \begin{pmatrix} S_1^x & S_1^y & S_1^z \\ S_2^x & S_2^y & S_2^z \\ S_3^x & S_3^y & S_3^z \end{pmatrix}
    \det \begin{pmatrix} S_1^x & S_2^x & S_3^x \\ S_1^y & S_2^y & S_3^y \\ S_1^z & S_2^z & S_3^z \end{pmatrix} \nonumber\\
    &= \det \begin{pmatrix} 1 & \VEC{S}_1\cdot\VEC{S}_2 & \VEC{S}_1\cdot\VEC{S}_3 \\
    \VEC{S}_2\cdot\VEC{S}_1 & 1 & \VEC{S}_2\cdot\VEC{S}_3 \\
    \VEC{S}_3\cdot\VEC{S}_1 & \VEC{S}_3\cdot\VEC{S}_2 & 1 \end{pmatrix} \nonumber\\
    &= 1 + 2 \left(\VEC{S}_1\cdot\VEC{S}_2\right) \left(\VEC{S}_2\cdot\VEC{S}_3\right) \left(\VEC{S}_3\cdot\VEC{S}_1\right) \nonumber\\
    &\hspace{1em}- \left(\VEC{S}_1\cdot\VEC{S}_2\right)^2 - \left(\VEC{S}_1\cdot\VEC{S}_3\right)^2 - \left(\VEC{S}_2\cdot\VEC{S}_3\right)^2 \;.
\end{align}
It is the sum of a constant, an isotropic 6-spin 3-site interaction, and three isotropic biquadratic interactions.
The isotropic 6-spin interaction written as sums over triples of dot products of spins was derived from a Hubbard model in Ref.~\onlinecite{MacDonald1990} (see last row of Table II in that work).

We now address the spin-chiral interaction\cite{Grytsiuk2020} (SCI) and its reduction to combinations of dot products and cross products.       
In three dimensions, any vector $\VEC{v}$ can be written using three linearly-independent vectors $\{\VEC{a},\VEC{b},\VEC{c}\}$ as
\begin{align}
    \VEC{v} =  \frac{\VEC{v} \cdot (\VEC{b} \times \VEC{c})}{\VEC{a} \cdot (\VEC{b} \times \VEC{c})} \, \VEC{a}
    + \frac{\VEC{v} \cdot (\VEC{c} \times \VEC{a})}{\VEC{a} \cdot (\VEC{b} \times \VEC{c})} \, \VEC{b}
    + \frac{\VEC{v} \cdot (\VEC{a} \times \VEC{b})}{\VEC{a} \cdot (\VEC{b} \times \VEC{c})} \, \VEC{c} \;. \label{eq:decomposition_3_vectors}
\end{align}
Using this relation, we can rewrite a generalized spin-chiral-type interaction as
\begin{align}
    \MC{E}_{1234}^\MR{SCI} &= \left(\VEC{C}_{1234}^\MR{SCI} \cdot \VEC{S}_1\right) \VEC{S}_2\cdot\left(\VEC{S}_3 \times \VEC{S}_4\right) \nonumber \\
    &= \VEC{C}_{1234}^\MR{SCI} \cdot \big( \left( \VEC{S}_3 \times \VEC{S}_4\right) \left(\VEC{S}_1 \cdot \VEC{S}_2\right)
    - \left( \VEC{S}_2 \times \VEC{S}_4 \right) \left(\VEC{S}_1 \cdot \VEC{S}_3\right) \nonumber \\
    &\hspace{4em}+  {\left( \VEC{S}_2 \times \VEC{S}_3 \right) \left(\VEC{S}_1 \cdot \VEC{S}_4\right)}\big) \;.
\end{align}
%The 4-site SCI has the six symmetries of the scalar spin chirality built from $234$, while the general chiral 4-site form given in Eq.~\eqref{eq:4site} has only four symmetries, so it cannot be mapped to the 4-site SCI.
The general chiral 4-site form given in Eq.~\eqref{eq:4site} has four symmetries while the 4-site SCI has the six symmetries of the scalar spin chirality built from $234$, so a mapping is not possible.
The 3-site restriction of this formula (the one actually discussed in Ref.~\onlinecite{Grytsiuk2020}) yields
\begin{align}
    \MC{E}_{1123}^\MR{SCI} &= \left(\VEC{C}_{1123}^\MR{SCI} \cdot \VEC{S}_1\right) \VEC{S}_1\cdot\left(\VEC{S}_2 \times \VEC{S}_3\right) \nonumber \\
    &= \VEC{C}_{1123}^\MR{SCI} \cdot \big(\VEC{S}_2 \times \VEC{S}_3
    +  \left( \VEC{S}_3 \times \VEC{S}_1 \right) \left(\VEC{S}_1 \cdot \VEC{S}_2\right) \nonumber \\
    &\hspace{5em}- {\left(\VEC{S}_2 \times \VEC{S}_1 \right) \left(\VEC{S}_1 \cdot \VEC{S}_3\right)}\big)   \;.
\end{align}
The first term contributes to the 2-spin DMI between sites 2 and 3, and the remaining two terms fall into the form of the 4-spin 3-site chiral interaction given in Eq.~\eqref{eq:3site}.
In fact, this corresponds to an antisymmetrized form of the chiral 3-site interaction, $\VEC{C}_{312}^- \cdot \big(\VEC{S}_3 \times \VEC{S}_1 \left(\VEC{S}_1 \cdot \VEC{S}_2\right) - {\VEC{S}_2 \times \VEC{S}_1 \left(\VEC{S}_1 \cdot \VEC{S}_3\right)}\big)$, so it cannot capture the symmetrized remainder of the chiral 3-site interaction.
%The 4-site SCI has the symmetry $\VEC{C}_{1123}^\mathrm{SCI} = -\VEC{C}_{1132}^\mathrm{SCI}$ which results in three independent terms instead of the six found for the general chiral 3-site form given in Eq.~\eqref{eq:4site}.

\section{Symmetry operations for $C_\MR{3v}$ and $C_\MR{4v}$}\label{app:symm}
For $C_\MR{3v}$ symmetry we give two rotation and three mirror matrices.
The $\SI{120}{\degree}$ rotations around the $z$-axis are
\begin{equation}
    \renewcommand{\arraystretch}{1.2}
    \MC{R}_\pm = \begin{pmatrix} -\frac{1}{2} & \mp\frac{\sqrt{3}}{2} & 0 \\
    \pm\frac{\sqrt{3}}{2} & -\frac{1}{2} & 0 \\ 0 & 0 & 1\end{pmatrix} \;.
\end{equation}
$\MC{R}_+$ being anticlockwise and $\MC{R}_-$ clockwise.
The matrices for the mirror planes in Fig.~\ref{fig:symmetries_illustration}a are
\begin{equation}
    \renewcommand{\arraystretch}{1.2}
    \MC{M}_1 = \begin{pmatrix} \frac{1}{2} & \frac{\sqrt{3}}{2} & 0 \\ \frac{\sqrt{3}}{2} & -\frac{1}{2} & 0 \\ 0 & 0 & 1 \end{pmatrix} \;,\quad
    \MC{M}_2 = \begin{pmatrix} \frac{1}{2} & -\frac{\sqrt{3}}{2} & 0 \\ -\frac{\sqrt{3}}{2} & -\frac{1}{2} & 0 \\ 0 & 0 & 1 \end{pmatrix} \;,
\end{equation}
and $\MC{M}_3 = \MR{diag}(-1,1,1)$ is a diagonal matrix.
    
For $C_\MR{4v}$ symmetry we define three rotation matrices and four mirror planes.
The reference rotation matrix is the anticlockwise rotation by $\SI{90}{\degree}$ around the $z$-axis,
\begin{equation}
    \renewcommand{\arraystretch}{1.2}
    \MC{R} = \begin{pmatrix} 0 & -1 & 0 \\ 1 & 0 & 0 \\ 0 & 0 & 1\end{pmatrix} \;,
\end{equation}
from which all rotation matrices can be defined by a suitable power or transpose.
We have two mirror planes along the cartesian axes,
\begin{equation}
    \MC{M}_x = \begin{pmatrix} -1 & 0 & 0 \\ 0 & 1 & 0 \\ 0 & 0 & 1 \end{pmatrix} \;, \quad
    \MC{M}_y = \begin{pmatrix} 1 & 0 & 0 \\ 0 & -1 & 0 \\ 0 & 0 & 1 \end{pmatrix} \;,
\end{equation}
and two diagonal ones
\begin{equation}
    \MC{M}_+ = \begin{pmatrix} 0 & 1 & 0 \\ 1 & 0 & 0 \\ 0 & 0 & 1 \end{pmatrix} \;, \quad
    \MC{M}_- = \begin{pmatrix} 0 & -1 & 0 \\ -1 & 0 & 0 \\ 0 & 0 & 1 \end{pmatrix} \;.
\end{equation}

\end{appendix}

\bibliography{Bibliography.bib}

\end{document}